\documentclass[12pt,preprint]{aastex}
\usepackage{graphics}

\begin{document}
\slugcomment{Submitted to ApJ supplements}

\title{CHANDRA ACIS SURVEY OF X-RAY POINT SOURCES IN 383 NEARBY GALAXIES I. THE SOURCE CATALOG}

\author{Jifeng Liu}
\affil{Harvard-Smithsonian center for Astrophysics, 60 Garden st. Cambridge, MA 02138}

\begin{abstract}

The Chandra data archive after eight years' accumulation is a treasure for
various studies, and in this paper we exploit this valuable resource to study
the X-ray point source populations in nearby galaxies.
By December 14, 2007, 383 galaxies within 40 Mpc with isophotal major axis
above 1 arcminute have been observed by 626 public ACIS observations, 60\% of
which are for the first time analyzed by this survey to study the X-ray point
sources.
Uniform data analysis procedures are applied to the 626 ACIS observations and
lead to the detection of 28099 point sources, which belong to 17599 independent
sources.
These include 8700 sources observed twice or more and 1000 sources observed 10
times or more, providing  us a wealth of data to study the long term
variability of these X-ray sources.
Cross correlation of these sources with galaxy isophotes led to 8519 sources
within the D25 isophotes of 351 galaxies, 3305 sources between the D25 and 2D25
isophotes of 309 galaxies, and additionally 5735 sources outside 2D25 isophotes
of galaxies.
This survey has produced a uniform catalog, by far the largest, of 11824 X-ray
point sources within 2D25 isophotes of 380 galaxies with 71\% of them reported
for the first time.
Contamination analysis using the logN-logS relation shows that 74\% of sources
within 2D25 isophotes above $10^{39}$ erg s$^{-1}$, 71\% of sources above
$10^{38}$ erg s$^{-1}$, 63\% of sources above $10^{37}$ erg s$^{-1}$, and 56\%
of all sources are truly associated with galaxies. 
This archival survey leads to 300 ULXs with 
$L_X(0.3-8{\rm keV})\ge2\times10^{39}$ erg s$^{-1}$
within D25 isophotes, 179 ULXs between D25 and 2D25 isophotes, and a total of
479 ULXs within 188 host galaxies, with about 324 ULXs truly associated with
host galaxies based on the contamination analysis.
About 4\% of the sources 
exhibited at least one supersoft phase, and 80 sources are classified as
ultraluminous supersoft sources with 
$L_X(0.3-8{\rm keV})\ge2\times10^{38}$ erg s$^{-1}$.
With a uniform dataset and good statistics, this survey enables future works on
various topics, such as 
X-ray luminosity functions, and multiwavelength identification and
classification.

\end{abstract}

\keywords{catalogs -- galaxies: general -- X-rays: binaries -- X-rays: galaxies}

\section{INTRODUCTION}

Starting from the historic discovery of Sco X-1 (Giacconi et al. 1962), X-ray
observations of more than forty years (see Pounds 2004 for a review) have
revealed a variety of X-ray point sources beyond the solar system in the Milky
Way and Magellanic Clouds, including low-mass X-ray binaries, high-mass X-ray
binaries, X-ray pulsars, cataclysmic variables, X-ray novae, and X-ray stars.
The most energetic X-ray emission comes from the accretion disks around compact
objects, follow-up observations of which 
can lead us to find the intriguing black holes and neutron stars out of
zillions of optical objects.
Indeed, follow-up optical observations have been able to measure the primary
masses for 20 X-ray binaries as above 3$M_\odot$, which places them securely as
black holes (McClintock \& Remiliard 2006).
Although the bright X-ray sources in the Milky Way are easily studied, many of
them are not observable due to the heavy obscuration of the galactic disk, and
one must correct for completeness when studying the statistical properties of
Galactic X-ray binaries (e.g., Grimm et al. 2002).
On the other hand, studies of X-ray binaries in more distant galaxies may be
free from the obscuration problem, and more importantly, provide us with
uniform samples of X-ray binaries in a variety of different environments.
It was, however, difficult to clearly resolve the point sources from the
diffuse emission due to hot gas in these distant galaxies with the limited
spatial resolution of X-ray observations prior to Chandra, and study of point
sources was only possible for the nearest galaxies (see Fabbiano 1989 for a
review).

With the unprecedented sub-arcsecond spatial resolution of Chandra, it is now
possible to separate even very closely spaced point sources and easily
distinguish them from surrounding diffuse emission, and reach much lower
sensitivity limits than was previously possible.
Much work has been done to study the X-ray source populations in distant
galaxies since the launch of Chandra (see Fabianno \& White 2006 for a review).
For example, the point source populations have been examined in great detail
and depth for starburst galaxies (e.g., the Antennae, Fabbiano et al. 2001),
spiral galaxies (e.g., M101, Kuntz 2003; M81, Pooley et al. 2004) and
early-type galaxies (e.g., NGC4697, Sarazin et al. 2001).
%
%
Kilgard et al. (2002) studied 3 starburst and 5 spiral galaxies, and found that
starburst galaxies have flatter X-ray point source luminosity function (XLF)
slopes than do normal spirals, as confirmed by a later study of 11 nearby
face-on spiral galaxies (Kilgard et al. 2005).
Grimm et al. (2003) studied the X-ray sources (mainly high-mass X-ray binaries)
in a dozen of late-type and starburst galaxies, and found that the total X-ray
luminosity and the XLF are scaled by the star formation rate.
New classes of X-ray sources not previously seen in the Milky Way have emerged
from studies of distant galaxies, such as the ultraluminous X-ray sources
(Miller \& Colbert 2004) and ultraluminous supersoft sources (Liu \& di Stefano 2008)
that are possibly the long sought intermediate mass black holes of a few
thousand solar masses.

Chandra Data Archive is a valuable resource for various studies on different
topics of X-ray astronomy. 
The Serendipitous Extragalactic X-ray Source Identification project (SEXSI;
Harrison et al. 2003) has used archival data from 27 Chandra fields covering
more than 2 deg$^2$ to study sources at intermediate fluxes and provided an
essential complement to the Chandra Deep fields in studying the extragalactic
hard X-ray sky. 
The Chandra Multiwavelength project (ChaMP; Kim et al. 2004) has analyzed 149
archival observations covering 10 deg$^2$ (Kim et al. 2007) in order to
investigate the formation and evolution of high-redshift AGNs, properties of
X-ray luminous galaxies and clusters, and constituents of cosmic X-ray
background.
The Chandra Multiwavelength Plane (ChaMPlane) survey utilizes the growing body
of Chandra archival observations and the NOAO 4 m Mosaic imaging survey to
measure or constrain the populations of low-luminosity accreting white dwarfs,
neutron stars, and stellar mass black holes in the Galactic plane and bulge
(Grindlay et al. 2005).
Colbert et al. (2004) studied 1441 X-ray sources in 32 nearby spiral and
elliptical galaxies and confirmed the correlation between the total point
source X-ray luminosity and both the old and young stellar populations.
Swartz et al. (2004) studied 154 ULXs in 82 nearby galaxies and confirmed the
preponderance of ULXs in star-forming galaxies as well as their similarities to
less-luminous sources, suggesting that ULXs originate in a young but
short-lived population.

Here we utilize the wealth of Chandra archival data from eight years'
accumulation to study the X-ray point source populations in 
nearby galaxies.
While Chandra studies of individual galaxies have led to many interesting
results,
these results need to be tested with large samples of galaxies, and this study
will enable such tests with good statistics. 
As the first in a series, we present in this paper the uniform procedures for
source detection and a catalog of extragalactic sources that can be used by
other researchers.
In section 2, we describe our selection of ACIS observations and nearby
galaxies that has resulted in 626 ACIS observations covering 383 galaxies
within 40 Mpc with isophotal major axis above 1 arcminute. 
In section 3, we describe the uniform procedures applied to all ACIS
observations to detect and visually check point sources, to compute the source
counts and colors, to check the source variability, to compute the source
spectrum and flux, and to assess the source position uncertainties.
In section 4, we describe how we cross identify sources in multiple
observations and compute upper limits for sources observed but not detected.
In section 5, we present the catalog for 11824 point sources in 380
nearby galaxies and the catalog for their individual observations.
We present the contamination analysis for this survey in section 6, and present
80 ultraluminous supersoft sources and 479 ultraluminous X-ray sources within
188 host galaxies in section 7.
We summarize our results and conclude with future works in section 8.

\section{SELECTION OF ACIS OBSERVATIONS AND NEARBY GALAXIES}

The nearby galaxy sample is extracted from the Third Reference Catalog of
Galaxies (RC3; de Vacouleous 1991), which is complete for nearby galaxies
having apparent diameters larger than 1 arcminute at the $D_{25}$ isophotal
level, total B-band magnitudes $B_T$ brighter than 15.5, and redshifts not in
excess of 15,000 km/sec.
The positions and the $D_{25}$ isophotes of the galaxies are taken from RC3.
These positions are usually better to $5^{\prime\prime}$, as verified by
comparing the surveyed galaxies to the DSS images. 
%


Distances to galaxies are collected from the literature, including the HST Key
Project (designated as KP; Freedman et al. 2002), the surface brightness
fluctuation method (SBF; Tonry et al. 2001), the nearby galaxy flow model by
Tully (1992; T92), and the Nearby Galaxy Catalog (T88; Tully 1988). Distances
are also computed using the Hubble relation $v = H_0D$ with $H_0=75$ km/s/Mpc.
The NED service provides the recessional velocity $v$ for most galaxies, and
distances computed from them are designated as NED. Another source of $v$ is
RC3, and the distances computed are designated as V3K, since we use the
velocities corrected to the rest frame defined by the 3K cosmic microwave
background.  When multiple distance measurements are available we use the best
distance measurement, which is KP, followed by SBF, T92, T88, NED, and V3K.


The flux densities of galaxies at 60$\mu$m are taken from the IRAS point source
catalog (IPAC, 1986), with some nearby galaxies from Rice et al. (1988).  When
the galaxy is below detection, the $3\sigma$ upper limit is calculated by
adopting noise levels of 8.5 mJy/arcmin$^2$ for 60 $\mu$m (Rice et al. 1988). 
Here we adopt such emission as an indicator for star formation rate (i.e.,
${\rm SFR}(M_\odot yr^{-1}) = 4.5\times10^{-44} L(60\mu m)$; Rosa-Gonzalez et
al. 2002), because such fluxes are available for most of the survey galaxies.
we note, however, IRAS does not have the resolution to excise the AGN, which can
contribute enormous FIR emission and lead to erroneous SFR estimates.


For studying extragalactic point sources, we choose galaxies with isophotal
diameters in excess of 1 arcminute and within 40 Mpc. We exclude M31, LMC and
SMC that are too large as compared to Chandra fields of view; we also exclude
M87 and close neighbors for its intricate jet and diffuse emission (Forman et
al. 2007).
These selection criteria lead to a RC3 galaxy sample of 4484 galaxies.
%
The list of ACIS archive observations is extracted from the Chandra Data
Archive\footnote{http://cxc.harvard.edu/cda/} on 
December 14, 2007, which includes 3004 ACIS-S observations and 1885 ACIS-I
observations.
Cross correlation of the ACIS observation list and the RC3 nearby galaxy
sample, with a radius of $5^{\prime}$ for ACIS-S observations and $9^\prime$
for ACIS-I observations, leads to an ACIS sample of 383 galaxies observed by
626 observations. In addition, 14 galaxies within 40 Mpc but with isophotal
diameters $<1$ arcminute are observed by these selected observations.

The survey galaxies, taking up 8.5\% of the 4484 RC3 galaxies within 40 Mpc,
are listed in Table 1, with galactic positions, sizes, the Galactic HI column
densities, the distances, the galaxy types, the B-band, the IRAS 60$\mu$m and
FIR luminosities.
The ACIS survey galaxy sample is compared to the RC3 galaxy sample to reveal
possible preferences in the survey.
Similar to the ROSAT HRI survey (Liu \& Bregman 2005), the  blue luminosity
distribution of the survey galaxies resembles that of the RC3 galaxy sample to
a large extent, as shown in Figure 1, except for a slight over-sampling of very
bright galaxies.
This over-sampling is related to the fact that the galaxies surveyed by the
Chandra ACIS tend to be larger and closer ones, as demonstrated in Figure 2 and
Figure 3.
The comparison between the morphological type distributions of the two galaxy
samples (Figure 4) shows an over-abundance of lenticulars and ellipticals, and
over-abundances of S0/a--Sb early spirals and peculiars to a lesser extent. In
contrast, there is an under-abundance of dwarf spirals and irregulars in the
survey sample.

The 626 ACIS observations selected for this survey include 83 ACIS-I
observations and 543 ACIS-S observations. 
For each observation, we use the on-axis chips, which include the S2 and S3
chips when the aimpoint is on S3, and all four I chips when the aimpoint is on
an I chip.
The exposure times for the selected observations range from 500 seconds to 120
kilo-seconds, with an average of 16 ksec and a total of 16,292 ksec.
Among these observations, more than half ($\ge$60\%) are designed to study
supernovae, jet, interstellar medium, superwinds, accretion flows, dark matter
halos, QSOs, AGNs, or galaxy group/clusters, and this survey is the first
effort to analyze the X-ray point source populations of those galaxies.

We have placed the 626 ACIS observations into 320 groups based on the proximity
of the pointings.
The observations in the same group have overlapping fields of view, and are
suitable to be studied together.
There are 48 groups with two ACIS observations, and 45 groups with three and
more observations. The latter includes the M33 group observed 28 times over six
years, the M101 group observed 26 times over five years, and the PGC3598 group
observed 21 times over 8 months. 
These multiple observations of the same sky region provide us the chance to
study the long term variability of the point sources.
We list in Table 2 for each group the observed galaxies within 40 Mpc, the
number of observations in the group, the total exposure time, and the
individual observations.
For each observation, we list the observation ID, observation date, the
modified Julian Date, the exposure time, the chips used for analysis, and the
total number of point sources detected and verified by visual inspection on
these chips with uniform procedures as described in \S3.

\section{ANALYSIS OF ACIS OBSERVATIONS}

For each ACIS observation, we apply the same procedures to detect and visually
check point sources, to compute the source counts and colors, to check the
source variability, to compute the source spectrum and flux, and to assess the
source position uncertainties.
The ACIS observations were downloaded from the Chandra Data Archive on December
14, 2007, and analyzed using CIAO 3.4.

\subsection{Source detection and visual examination}

For point source detection, we choose {\tt wavdetect} that adopts a wavelet
algorithm and is largely used for Chandra observations (Freeman et al. 2003).
This tool is able to separate closely spaced point sources, and able to
recognize diffuse emission quite well for its advanced treatment of the
background.  
We run {\tt wavdetect} on each on-axis chip with scales of $1^{\prime\prime},
2^{\prime\prime}, 4^{\prime\prime}$, and $8^{\prime\prime}$ in the 0.3-8 keV
band. The chance detection threshold is set to $10^{-6}$, equivalent to 1
spurious detection per chip. 

The sources reported by {\tt wavdetect} are visually examined for possible
false detections.
Spurious point sources could emerge on the CCD edges, from streaks, or from the
Lissajous patterns on S2 due to the combination of telescope wobble and a
bright source falling in a bad pixel.
Such sources, and sources with zero sizes, are removed from the source lists
after visually examining the detected sources overlayed on the X-ray images.
Despite the capacity of {\tt wavdetect} to recognize diffuse emission, spurious
point sources may arise from diffuse emission with very complex morphologies,
e.g., those from the X-ray jets (such as in NGC5128) and starburst regions.
Such sources are also identified and removed with visual examination.
Occasionally, {\tt wavdetect} may split a point source at large off-axis angles
with distorted PSF shapes into two sources. We identify such pairs of split
sources and merge them as single sources.
In the end, we have detected and verified 28099 point sources from the 1355 chips of the 626
observations in this survey.
We note the crucial role of visual inspection in this work, which has removed
about 10\% of the total number of sources originally reported by wavdetect.

\subsection{Source counts}

The photon counts of sources are derived with two methods.
{\tt wavdetect} computes the source counts for detected sources by fitting a
2-D Gaussian to the image.
Alternatively, we perform aperture photometry to directly count the photons
within the source region with expected background counts subtracted. 
In this work, the source region is taken as the $3\sigma$ elliptical source
region as computed by {\tt wavdetect}, which contains 95\% of the source count
for an assumed 2-D Gaussian distribution.
For the background region, we use an elliptical annulus around the source
region with the outer/inner radius as 4/2 times the radius of the source
ellipse.  The background regions are checked to exclude any other source
ellipses within them. 
The source count is computed as $C_s - C_b\times A_s/A_b$, and the error is
computed as $1 + \sqrt{0.75 + C_s + C_b \times (A_s/A_b)^2}$, with $C_s$ as the
raw source count, $A_s$ as the source region area, $C_b$ as the background
count, and $A_b$ as the background region area.

The source counts obtained from the two methods are compared for all sources
detected in these ACIS observations as shown in Figure 5.
For sources with detection significance $\sigma>20$, the source counts from
aperture photometry are consistent with 95\% of the wavdetect photon counts,
with 1-$\sigma$ dispersions of 2\%-8\%.
At lower detection significance, the source counts from aperture photometry can
deviate from the wavdetect counts by up to 50\%, but the average counts deviate
by no more than 5\% unless for those marginal detections with $\sigma<3$.
The wavdetect counts are used to compute the fluxes and luminosities in the
0.3-8 keV band.

\subsection{Source colors}

The photon counts are computed for different energy bands to compute X-ray
colors for the sources. 
The combination of the X-ray colors may distinguish between AGNs, supernovae,
stars, LMXBs and HMXBs because they occupy different locations in the
color-color diagram (e.g., Prestwich et al. 2003).
We compute the photon counts with aperture photometry for three bands following
Prestwich et al. (2003), i.e., the soft band (S:0.3-1.0 keV), the medium band
(M:1.0-2.0 keV) and the hard band (H:2.0-8.0 keV).
The X-ray colors are computed as $C_{MS} = (M-S)/(H+M+S)$ and $C_{HM} =
(H-M)/(H+M+S)$.
Note that the band counts can be slightly less than 0 due to over-subtraction
of the background.
Figure 6 shows all detected sources above 10 counts from this survey in the
color-color diagram. About 2\% falls into the SNR region, 11\% falls into the
HMXB region, 42\% falls into the LMXB region, and 43\% falls into the region
for absorbed sources that are probably a mixture of types.
We caution that these regions, based on a few hundred sources in five nearby
galaxies (Prestwich et al. 2003), may not correctly categorize the X-ray
sources. For example, some sources in the SNR region show clear variability
within one observation and/or between different observations, thus are certainly
not SNRs.


Very soft X-ray sources, including supersoft sources (SSS) and quasi-soft
sources (QSS), are very common in external galaxies (Di Stefano \& Kong 2004). 
While some soft sources are possibly white dwarfs with steady nuclear burning
on their surface, some can be stripped stars, stellar black holes, or even
intermediate-mass black holes of a few $\times 10^3 M_\odot$ (e.g., M81-ULS1,
Liu \& Di Stefano 2008).
In this work, we compute the photon counts in three energy bands (0.1-1.1 keV,
1.1-2 keV, and 2-7 keV), and follow the hierarchical classification scheme (Di
Stefano \& Kong 2003) to check whether a source is SSS, QSS, hard, or too dim
(i.e., less than 10 counts).
About 24\% of all 28099 detected sources are too dim, among the rest about
2.6\% are classified as SSS, 10.3\% as QSS, and 87.1\% as hard. On the X-ray
color-color diagram as in Figure 6, QSS sources are located along H=0, and SSS
sources are clustered around $C_{MS} = -1, C_{HM} = 0$.

\subsection{Source Variability}

The Chandra observations are usually consecutive with occasional gaps of a few
seconds, ideal for testing short timescale variability of the sources.
The event list of a source is extracted from the $3\sigma$ elliptical source
region.
The nonparametric Kolmogorov-Smirnov test is applied to the event list to test
the null hypothesis that the source is constant during the observation.
A source can be viewed as constant if the null hypothesis probability $P_{KS}$
is of the order of 1.
A source can be viewed as variable if the probability is much smaller, for
example, $\le$0.1. To be conservative, we define a source as variable if
$P_{KS}<0.01$, i.e., the source is variable with a significance of $>$99\%.
By this criterion, about 4\% (922/23957) of the detected sources above 10 counts are
variable within one observation.

The binned light curve is constructed for a source to visualize its variability
within an observation.
An example is shown in Figure 7, which exhibits a possible X-ray eclipse.
A closer look at this source is thus solicited and, with 25 ACIS observations,
has led to the discovery of an X-ray eclipsing binary in M101 with an orbital
period of $32.688 \pm 0.002$ hr, among the first of such outside the Local
Group (Liu et al. 2006).
We calculate the reduced $\chi^2 = \sum_{i=1}^N
(C_i-\bar{C})^2/[(N-1)\bar{C}^2]$, with the light curve binned to $N$ bins so
that the average count per bin $\bar{C}$ is no less than 10.
The reduced $\sqrt{\chi^2}$ can serve as a measure of the variation amplitude
(Collura et al.  1987).
For about 10,000 sources above 40 counts, the reduced $\sqrt{\chi^2}$ ranges
from 0.04 (steady) to 10 (extremely variable).
The average for these sources is 0.32, and is 0.61 for sources variable by the
K-S criterion.

\subsection{Source spectrum and flux}

For a bright point source, we extracted the spectrum from the $3\sigma$
elliptical source region with {\tt psextract}. 
This tool automatically accounts for the spatial and time dependence of the
ACIS quantum efficiency due to the buildup of contamination on the optical
blocking filter.
The corresponding background spectrum for a source is extracted from its local
background region as specified in \S3.2.
An absorbed power-law model is fitted to each spectrum; we set the Galactic
$n_H$ as the minimum absorption column density.
Most (95\%) of the spectra can be fitted by the absorbed power-law model with
the photon index between 1 and 4. As shown in Figure 8, the photon index
distribution peaks at $\sim1.8$, with 68.3\% enclosed between 1.3 and 2.3.

The source flux is computed for each source from its count rate in the 0.3-8
keV band.
While the source flux can be accurately computed from the spectrum for some
bright sources, the flux must be derived from its count rate for most sources
without a spectrum.
The count rate is computed from the source count and the exposure time
corrected by a vignetting factor, which is derived from the exposure map as the
ratio between the local and the maximum map value.
The conversion factor between the count rate and flux is computed in {\tt
xspec} using the response matrix of the chip center, assuming Galactic
absorption and a power-law spectrum with a photon index $\Gamma=1.7$ as
conventionally adopted for Chandra sources.

Swartz et al. (2004) fitted power-law model to the ULX candidates from their
Chandra survey of nearby galaxies, and found 80\% of them have an average
photon index of 1.74, while the remaining 20\% are much steeper. 
If the source actually has a steeper spectrum, the true flux will be lower than
calculated with $\Gamma=1.7$ by, e.g., 25\% if $\Gamma=2.4$.
If the source has a harder AGN-like spectrum with $\Gamma=1.4$, the true flux
will be higher by 20\%.
The flux can be seriously under estimated if the source is highly absorbed as
in some AGNs. For example, the flux will be under-estimated by a factor of two
if the absorption is $n_H = 4\times10^{21}$ instead of the Galactic-like
$2\times10^{20}$ cm$^{-2}$.
On the other hand, if the underlying spectrum is a thermal one as in QSS/SSS,
the flux will be over estimated by 68\%/46\%/23\% for a blackbody of 350/150/70
eV with $n_H = 2\times10^{20}$ cm$^{-2}$.

\subsection{Source position uncertainties}

With the superb Chandra spatial resolution, the on-axis positional accuracy is
expected to be accurate with an error less than $1^{\prime\prime}$.
The positional accuracy becomes worse for faint sources at large off-axis
angles.
Simulations by Kim et al. (2004) show that the source position is accurately
determined for strong sources regardless of the off-axis angle $D_{off-axis}$.
For fainter sources, the positional error remains relatively small (less than
$1^{\prime\prime}-2^{\prime\prime}$) within $D_{off-axis} < 6^\prime$, and
increases to $2^{\prime\prime}-3^{\prime\prime}$ and
$4^{\prime\prime}-5^{\prime\prime}$ at $D_{off-axis} = 6^\prime-8^\prime$ and
$8^\prime-10^\prime$, respectively (95\% confidence).
Kim et al. (2004) gave a set of empirical formulas to approximately estimate
the positional error as a function of off-axis angle for a source of 20 counts
and a source of 100 counts. 
Here we adopt their scheme to compute the positional errors for a given
off-axis angle for 20 counts and 100 counts, and other counts by interpolations
and extrapolations.
As shown in Figure 9, thus computed positional errors are mostly (87\%) less
than $2^{\prime\prime}$, with  only 1\% larger than $5^{\prime\prime}$.
We apply a conservative minimum of $1^{\prime\prime}$ 
for these sources.
Note that the (absolute) positional error computed here is different from the
statistical positional uncertainty reported by {\tt wavdetect}, which is
usually small and meaningful for relative astrometric correction (e.g., Liu \&
Bregman 2001).


In summary, the above procedures lead to the detection of 28099 point sources
on 1355 on-axis chips for 626 ACIS observations. 
The number of detected sources for each observation is listed in Table 2. 
Of these sources, there are 734 (2.6\%) with detection significance $\sigma<2$,
3639 (13.0\%) with detection significance $\sigma<3$, and 13364 (47.6\%) with
$\sigma<6$ as shown in Figure 10.
The source distributions peak around 4.5$\sigma$, 14 counts after background
subtraction, and $5\times10^{-4}$ count/sec after vignetting correction (or
equivalently $3\times10^{-15}$ erg s$^{-1}$ cm$^{-2}$ for a power law spectral
model with $\Gamma=1.7$ and $n_H = 2\times10^{20}$ cm$^{-2}$).
This composite survey thus covers a dynamical range of 5 orders of magnitudes
in the source count (from 2 to $10^5$), the count rate (from $3\times10^{-5}$
to 3 count/sec) and the flux (from $2\times10^{-16}$ to $2\times10^{-11}$ erg
s$^{-1}$ cm$^{-2}$).
Our analysis with the above procedures results in uniform data products and
parameters for these sources.
%

%

\section{SOURCES IN MULTIPLE OBSERVATIONS}

To determine which source/detections in multiple observations belong to same
sources, we cross correlate the $3\sigma$ source ellipses as described in \S3.2
from different observations in the same group, and identify the correlated
source/detections as the same sources.
As shown in Figure 11, source/detections with overlapping source ellipses are
identified as same sources.
This straightforward method works quite well for most sources because they are
well separated due to the superb Chandra spatial resolution. 
Occasionally source ellipses from two nearby sources may overlap, as
illustrated by the two sources on the left of Figure 11, but we can remove such
confusions in most cases if we shrink the source ellipses to about the PSF
sizes.
In rare cases of extreme confusion, human judgment is used to determine the
sources.
Visual inspections are invoked to verify the cross identification of such
confusing situations.
In some observations, two close sources at large off-axis angles cannot be
resolved and may be detected as a single source. In this situation, the flux of
the detected source will be split into two components, and the relative
fractions are based on its separations from the centers of the two sources
determined from other observations.

Once the individual detections are determined for a source, the final position
is derived from averaging the positions of individual detections weighted by
the detection significance.
The minimum positional error from the individual detections is adopted as the
positional error.
If a source was observed but not detected by {\tt wavdetect} in an observation,
the upper limit is computed as the background-subtracted photon counts within
the PSF that encloses 95\% of the energy at 1.5 keV, with the background
computed from a surrounding annulus clear of detected sources. 
The source count is computed as $C_N = C_s - C_b\times A_s/A_b$, and the error
is computed as $C_E = 1 + \sqrt{0.75 + C_s + C_b \times (A_s/A_b)^2}$, with
$C_s$ as the raw source count, $A_s$ as the source region area, $C_b$ as the
background count, and $A_b$ as the background region area.
A minimum of 1 photon is set for $C_N$ if it is less than unit.
We define the source significance as $\sigma = C_N / C_E$, which is different
from the detection significance as reported by {\tt wavdetect} were it
detected.
Thus computed photon counts range from a few up to hundreds depending on the
off-axis angles and the environments of the sources.

The same cross identification method is applied to the 92 groups with two or
more observations, and lead to 10693 independent sources from 21800 detections. 
Among these sources, an average source is observed 4.3 times, and detected 2.0
times.
About 8700 sources were observed twice or more, with 1000 sources observed 10
times or more. 
This survey thus provide us a wealth of data to study the long term variability
of X-ray sources.
For each source, we compute $F_{max}$ as the maximum 0.3-8 keV flux from all
detections, $F_{min}$ as the minimum flux from all detections and upper limits,
and the $F_{max}/F_{min}$ ratio as an extreme indicator of its variability.
A more complete view of the long-term variability can be obtained from its
light curve, as shown in Figure 12 for an example.
For 7784 sources with different $F_{min}$ and  $F_{max}$ from this survey,
there are 109 sources (1.4\%) with $F_{max}/F_{min} \ge 100$, 1565 sources
(20\%) with $F_{max}/F_{min} \ge 10$, 2921 sources (37.5\%) with
$F_{max}/F_{min} \ge 5$, and 5617 sources (67\%) with $F_{max}/F_{min} \ge 2$.
In comparison, the variability during an observation, as revealed by the K-S
criterion, is exhibited in about 4\% (922/23957) of the detected sources above
10 counts, and in about 5.6\% (781/13755) of all independent sources with 10
counts or more in at least one observation.

\section{A CATALOG OF EXTRAGALACTIC X-RAY SOURCES}

The analysis of 320 groups of 626 ACIS observations leads to 17559 independent
sources from 28666 detections in this work.
To check whether a source is associated with a galaxy, we compute the
separation $\alpha$ between the galaxy center and the source and compare to the
elliptical radius $R_{25}$ of the $D_{25}$ isophotal ellipse along the great
arc connecting the galaxy center and the source. 
A source is considered as associated with the galaxy if $\alpha<R_{25}$, i.e.,
within the $D_{25}$ isophote.
This is because $D_{25}$ isophotes are a good delimiter of the optical domain
of the galaxies, though galactic features extend apparently beyond $D_{25}$,
but within 2 $\times D_{25}$ of some galaxies.
To not miss any galactic sources, a source with $R_{25} < \alpha < 2\times
R_{25}$ is tentatively associated with the galaxy.
In this survey, there are 8519 sources (including 7457 with detection
significance above 3$\sigma$) within D25 of 351 galaxies,  3305 sources
(including 2809 with detection significance above 3$\sigma$) between D25 and
2D25 of 309 galaxies, and a total of 11824 sources with 2D25 isophotes of 380
galaxies. In addition, there are 5735 sources outside 2D25 isophotes of
galaxies.
%


These extragalactic sources are listed in Table 3, ordered by group and
position. 
For each source, we list the group number, the CXO name composed from its
position, (absolute) positional error, galactic source name, nuclear separation
in arcminutes and in $R_{25}$, the distance, the number of
observations/detections, the maximum luminosity in 0.3-8 keV, the average flux,
the  $F_{max}/F_{min}$ ratio, the (maximum) detection sigma, the (maximum)
photon counts, the statistics for SSS/QSS and for variability in individual
observations, and the identifications of the source.
The identification indicates whether the X-ray source is an ultraluminous X-ray
source (ULX) or an ultraluminous supersoft source (ULS) as described below, or
identified with the galactic nucleus as detailed in a subsequent paper (Paper
II), or whether it was observed/cataloged previously in other wavelengths as
detailed in another paper (Paper IV).
In each galaxy, we number the X-ray sources based on their maximum detection
significance.
The 5735 sources outside 2D25 of galaxies are also included in Table 3. For
these sources, the luminosities are computed as if in a galaxy of that group;
they are not the true luminosities but listed for comparison.

The individual observations for these sources were listed in Table 4 with their
key parameters.
These parameters include the source number in that observation (and the split
fraction), Modified Julian Date of the observation, the exposure time, off-axis
angle, vignetting factor, statistical position uncertainty, absolute position
error, detection significance, source counts and (Gheral) errors for 0.3-8 KeV,
two colors for the 0.3-1 keV, 1-2 keV and 2-8 keV band set, flux for an assumed
$\Gamma=1.7$ power law, K-S test probability for constancy, and whether the
source is a SSS or QSS source.
If a source is a split component of a merged source in an observation as
described in \S4, the split fraction will be less than unit, and the listed
count and flux should be multiplied by this fraction to get its true count and
flux in this observation.
If a source was not detected in an observation and its upper limit was computed
as described in \S4, we prefix the computed source significance with a negative
sign, and list the background-subtracted photon count and error within its PSF
and the corresponding flux.

\section{CONTAMINATIONS OF THE CATALOG}

A key issue for extragalactic X-ray point source catalogs is how many of these
sources are actually foreground or background objects instead of objects truly
associated with the studied galaxies.
This contamination rate can be estimated with the logN-logS relation derived
from fields without nearby galaxies.
In this study, we adopt the logN-logS relation derived from ROSAT observations
(Hasinger et al. 1998) complemented at low fluxes by the logN-logS relation
derived from Chandra ACIS observations (Mushotzky et al. 2000).

Here we present the contamination estimates for the whole catalog, but
leave the technical details and more thorough analysis to a subsequent paper
(Paper III).
For each galaxy, we compute the survey area curve $A(>S)$ with the longest
observation that gives the area in which sources brighter than the limiting
flux $S$ can be detected in the observation.
The limiting flux is computed from the ($3\sigma$) detection threshold as a
function of off-axis angle and background rate, which we derive from a
collection of about 1600 sources detected at 2.8-3.2$\sigma$ in this survey.
The number of contaminating sources in a flux interval can then be calculated
with the differential form of the logN-logS relation.
For each galaxy, we have calculated the surveyed area curve, the (cumulative)
number of sources detected above $3\sigma$, the numbers of  the predicted
foreground/background sources, and the number of ``net'' sources truly
associated with the studied galaxy.

Estimates for individual galaxies are summed up to estimate the contamination
rate for the whole catalog.
The surveyed area curves for such a total survey are plotted in Figure 13. The
D25 isophotes of these 439 surveyed galaxies cover about 2.1 deg$^2$ of the
sky, while the 2$\times$D25 isophotes cover about 4.8 deg$^2$ of the sky.
In Figure 14 we plot the cumulative numbers of detected sources, predicted
contaminating sources and the ``net'' sources.
For all detected sources within the D25 isophotes, about 69.1\% are ``net''
sources truly associated with the studied galaxies, and 30.9\% are
foreground/background objects.
At larger luminosities, the net source fraction increases to 77.8\% for
detected sources above $10^{37}$ erg s$^{-1}$, to 85.4\% for detected sources
above $10^{38}$ erg s$^{-1}$, and to 87.9\% for detected sources above
$10^{39}$ erg s$^{-1}$.
For all detected sources between D25 and 2D25 isophotes, the total ``net''
source fraction is about 21\%, while the ``net'' source fraction increases to 25\% for
sources above $10^{37}$ erg s$^{-1}$, to 35\% for sources above $10^{38}$ erg
s$^{-1}$, and to 39\% for sources above $10^{39}$ erg s$^{-1}$.
Considering sources within 2D25 isophotes, the total ``net'' source fraction is
about 56\% for all 11824 sources, and increases to 63\%, 71\%, and 74\% for
sources above $10^{37}$, $10^{38}$, and $10^{39}$ erg s$^{-1}$.

In some sense, the nuclear X-ray sources, presumably powered by Bondi accretion
onto supermassive black holes in the galactic centers, are contaminations for
our study of the ordinary X-ray binary populations.
These nuclear X-ray sources are identified by comparing X-ray images with
optical images, and the details are reported in a subsequent paper (paper II). 
To summarize, 234 out of 390 galactic nuclei observed in this survey  are
identified to X-ray sources with luminosities ranging from $2\times10^{36}$ to
$2\times10^{42}$ erg s$^{-1}$.
Figure 14 shows the luminosity function for these nuclear sources, which is
much flatter and extends to much higher energies as compared to the luminosity
function for extranuclear sources in these galaxies.
For all detected sources within D25 isophotes, the nuclear sources take a
fraction of 4.9\%. This fraction increases to 6.6\% for X-ray sources above
$10^{38}$ erg s$^{-1}$, 
to 24\% for X-ray sources above $10^{39}$ erg s$^{-1}$, 
and to 67\% for X-ray sources above $10^{40}$ erg s$^{-1}$.
At X-ray luminosities of $10^{41}$ erg s$^{-1}$ and above, all detected sources
are nuclear sources in a total survey as shown in Figure 13.


\section{ULTRALUMINOUS X-RAY SOURCES}


This survey provides a valuable opportunity to study the statistics for
different X-ray source populations, including ultraluminous X-ray sources that
are possibly the long sought intermediate mass black holes of a few thousand
solar masses (Miller \& Colbert 2004). 
Here ultraluminous X-ray sources are defined as those nonnuclear sources with
maximum $L_{X}$(0.3-8keV) above $2\times10^{39}$ erg s$^{-1}$.
These sources are further classified as '1ULX' if they are within the D25
isophotes of the host galaxies, and as '2ULX' if they are outside the D25
isophotes but within 2D25 isophotes of the host galaxies.
In addition, extreme ULXs with maximum $L_{X}$(0.3-8keV) above $10^{40}$ erg
s$^{-1}$ are designated as 'EULX'.
These definitions lead to 300 1ULXs (including 47 1EULXs), 179 2ULXs (including
27 2EULXs) and a total of 479 ULXs within 188 host galaxies.
These ULXs are extracted from Table 3 and listed separately in Table 5 for the
readers' convenience.


The contamination rates for the ULX samples can be estimated with the logN-logS
relation.
For a total survey as described in \S6, 360 sources with $L_{X}$(0.3-8keV)
above $2\times10^{39}$ erg s$^{-1}$ are detected within the D25 isophotes of
the surveyed galaxies, including 107 galactic nuclei, about 40
foreground/background objects, and 320 ``net'' nonnuclear sources.
Since 253 (360-107) of these sources would be classified as 1ULXs, the net
source fraction of the 1ULX sample is about 84\% (213/253).
Similarly, the net source fraction is about 89\% for the 1EULX sample, about
40\% for the 2ULX sample, and about 41\% for the 2EULX sample.
Applying the above net source fractions to the ULX samples, we expect 252 true
1ULXs (including 42 true 1EULXs), 72 true 2ULXs (including 11 true 2EULXs), and
a total of 324 true ULXs. 


One special subclass of ULXs are the ultraluminous supersoft sources (ULSs).
Unlike most ULXs with a usually dominant hard power-law component, ULSs have a
pure thermal spectrum of temperature $\ll$ 1keV, which justifies linking the
emission to the accretion disk around an intermediate mass black hole (Liu \&
Di Stefano 2008).
The pure thermal spectrum can also come from the steady nuclear burning on the
surface of a white dwarf like the classical supersoft sources in the Milky Way
and Magellanic Clouds (Kahabka \& van den Heuvel 1997 and reference therein),
but the bolometric luminosity would be lower than the Eddington luminosity for
a white dwarf ($\approx2\times10^{38}$ erg s$^{-1}$).
For $\sim$12500 sources with 10 counts or more in at least one observation,
about 4\% exhibited as supersoft sources at least once, and 13\% as quasi-soft
sources at least once.
Here we define ULSs as those with at least one supersoft phase in which the
X-ray luminosity in 0.3-8 keV exceeds $2\times10^{38}$ erg s$^{-1}$. 
To summarize, this survey reveals 53 ULSs within the D25 isophotes of 36
galaxies, and 27 ULSs between the D25 and 2D25 isophotes of 18 galaxies.

\section{SUMMARY AND FUTURE WORKS}

The Chandra data archive after eight years' accumulation is a treasure for
various studies, and in this paper we exploit this valuable resource to study
the X-ray point source populations in nearby galaxies.
Cross correlation of the Chandra observation log with the nearby galaxy list
resulted in 626 ACIS observations covering 383 galaxies with isophotal major
axis in excess of 1 arcminute within 40 Mpc.
Among these observations, about 60\% were designed to study supernovae, jet,
interstellar medium, superwinds, accretion flows, dark matter halos, QSOs,
AGNs, or galaxy groups/clusters; this survey is the first effort to analyze the
X-ray point source populations of those galaxies.
Uniform data reduction and analysis procedures were applied to all 626 ACIS
observations to detect and visually check point sources, to compute the source
counts and colors, to check the source variability, to compute the source
spectrum and flux, and to assess the source position uncertainties. 

The uniform data analysis procedures detected 28099 point sources on 1355
on-axis chips for 626 ACIS observations.
%
These detections include 8700 sources observed twice or more and 1000 sources
observed 10 times or more, providing  us a wealth of data to study the long
term variability of these X-ray sources.
Cross correlation of these sources with galaxy isophotes led to 8519 sources
within the D25 isophotes of 351 galaxies, 3305 sources between the D25 and 2D25
isophotes of 309 galaxies, and additionally 5735 sources outside 2D25 isophotes
of galaxies.
This catalog was cross correlated with X-ray point source catalogs available in
the VizieR service, including 22 Chandra source catalogs, 7 XMM source catalogs
and 5 ROSAT source catalogs;  4378 sources (including 3410 within 2D25
isophotes) have matches within their positional errors.
In summary, this survey has produced a uniform catalog of 11824 sources within
2D25 isophotes of 380 galaxies with 71\% of them reported for the first time.
This is by far the largest catalog of X-ray point sources in nearby galaxies.

Some statistical properties for these extragalactic X-ray sources have been
derived.
Contamination analysis using the logN-logS relation shows that 74\% of sources
within 2D25 isophotes above $10^{39}$ erg s$^{-1}$, 71\% of sources above
$10^{38}$ erg s$^{-1}$, 63\% of sources above $10^{37}$ erg s$^{-1}$, and 56\%
of all sources are truly associated with galaxies. 
This archival survey leads to 300 ULXs with 
$L_X(0.3-8{\rm keV})\ge2\times10^{39}$ erg s$^{-1}$
within D25 isophotes, 179 ULXs between D25 and 2D25 isophotes, and a total of
479 ULXs within 188 host galaxies, with about 324 ULXs truly associated with
host galaxies based on the contamination analysis.
About 4\% of the sources 
exhibited at least one supersoft phase, and 80 sources are classified as
ultraluminous supersoft sources with 
$L_X(0.3-8{\rm keV})\ge2\times10^{38}$ erg s$^{-1}$.
Most X-ray sources are variable between observations separated by days to
years, and the extreme variability indicator $F_{max}/F_{min}\ge2$ for 67\% of
the sources with two or more observations.

This archival survey enables studies of many aspects of the X-ray source
populations in nearby galaxies, and a number of future works are planned to
better utilize this uniform dataset and its good statistics.
%
%
Firstly, to study the ordinary X-ray binary populations, we need to excise from
this catalog the nuclear X-ray sources that are presumably powered by Bondi
accretion of interstellar medium onto the supermassive black holes in the
galactic centers.
The nuclear X-ray sources observed in this survey are identified by comparing
X-ray images with optical images, and the details will be reported in a
subsequent paper (paper II). 
%
%
Secondly, while previous studies agree in the general picture that the X-ray
luminosity functions (XLFs) are flatter for starbursting galaxies than for
spiral or early-type galaxies, disagreements exist for some important details
such as the flattening of XLF at luminosities around a few $\times10^{37}$ erg
s$^{-1}$ and the break of XLF around $10^{39}$ erg s$^{-1}$.
In a subsequent paper (Paper III), we utilize this survey to construct a
library of XLFs for 380 galaxies of different types with uniform procedures,
and test the XLF evolution and details with good statistics.
%
%
Thirdly, to understand the nature of these X-ray sources, we need to identify
and study them in other wavelengths in addition to their X-ray spectral and
timing properties.
In another subsequent paper (Paper IV), we study the X-ray source
identification and classification in the optical, infrared, and radio bands
with 120 catalogs available in the VizieR service, and present the X-ray
properties for different classes derived from multiwavelength studies.


\acknowledgements

We would like to thank xxxx for helpful discussions.  JFL acknowledges the
support for this work provided by NASA through the Chandra Fellowship Program,
grant PF6-70043, and supports from grants blah.



\begin{figure}
\plotone{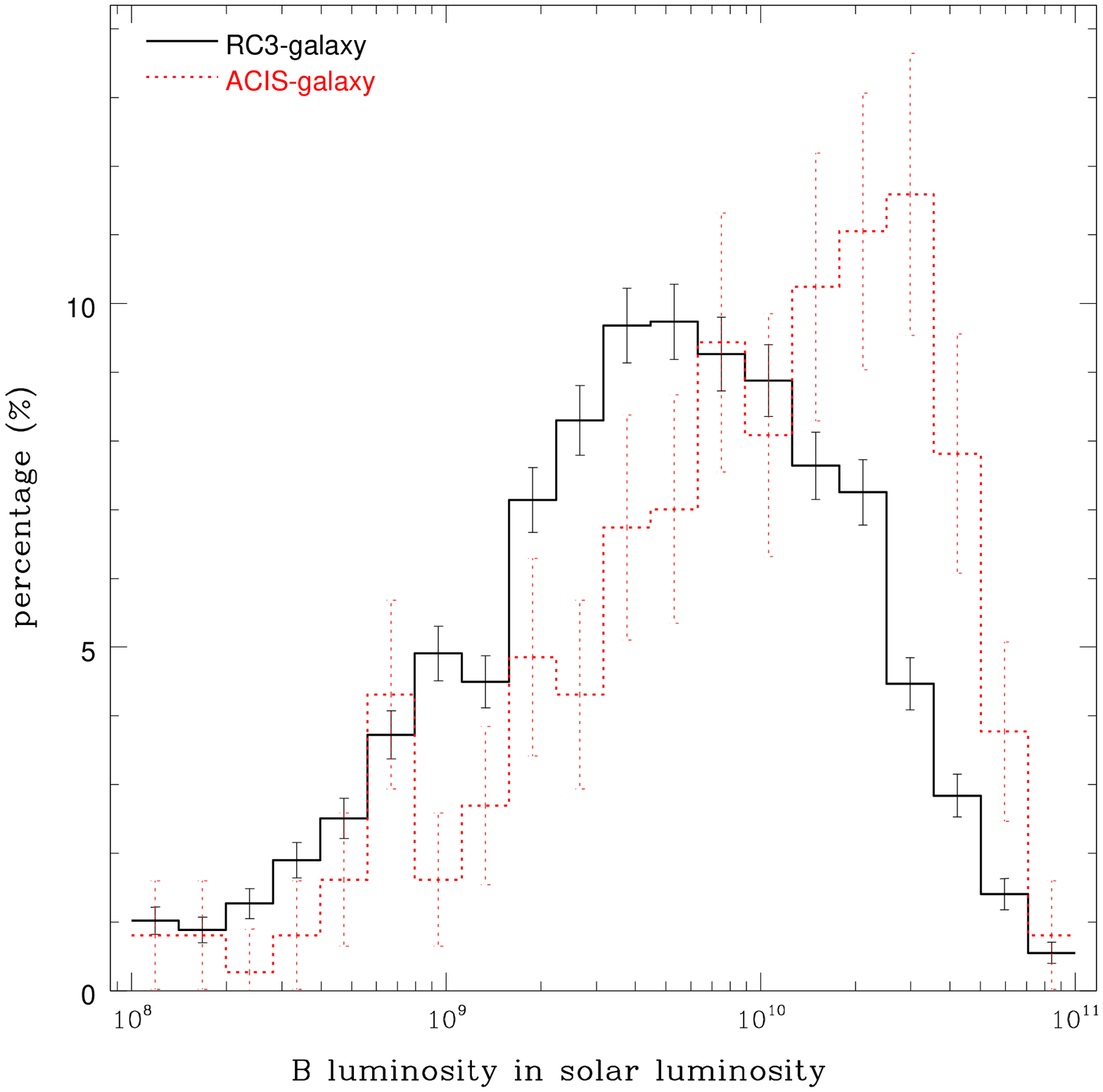}
\caption[Blue luminosities for ACIS survey galaxies]{The distribution of blue
luminosities for the Chandra ACIS survey galaxy sample in comparison with the
RC3 galaxy sample. More bright galaxies, presumably larger in size and closer
in distance, were surveyed by Chandra/ACIS than faint galaxies. }
\end{figure}


\begin{figure}
\plotone{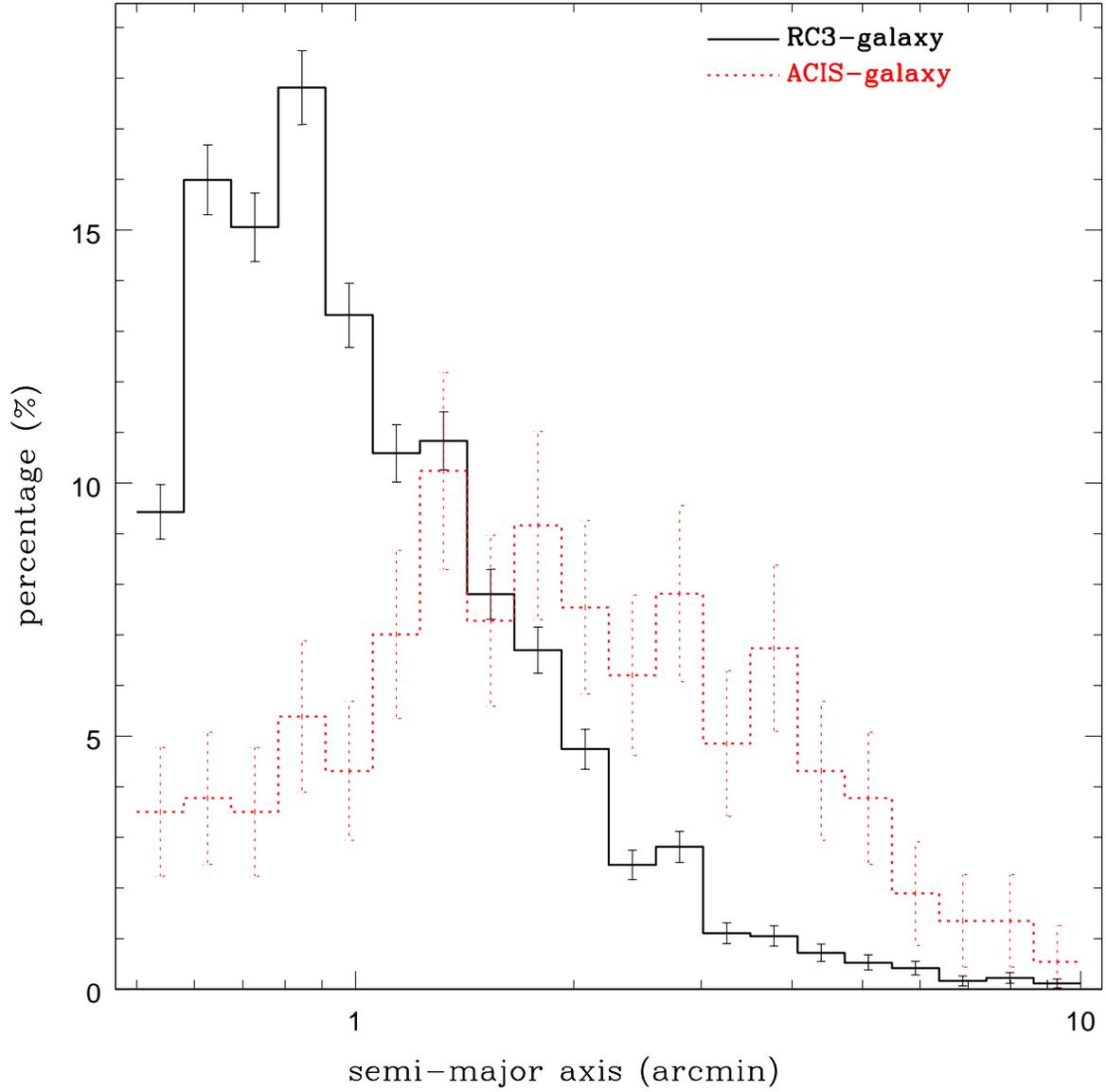}
\caption[Sizes for ACIS survey galaxies]{Galaxy size distribution of the
Chandra ACIS  survey galaxy sample in comparison with the RC3 galaxy sample.
Slightly more large galaxies were surveyed by Chandra/ACIS than small
galaxies.}

\end{figure}


\begin{figure}
\plotone{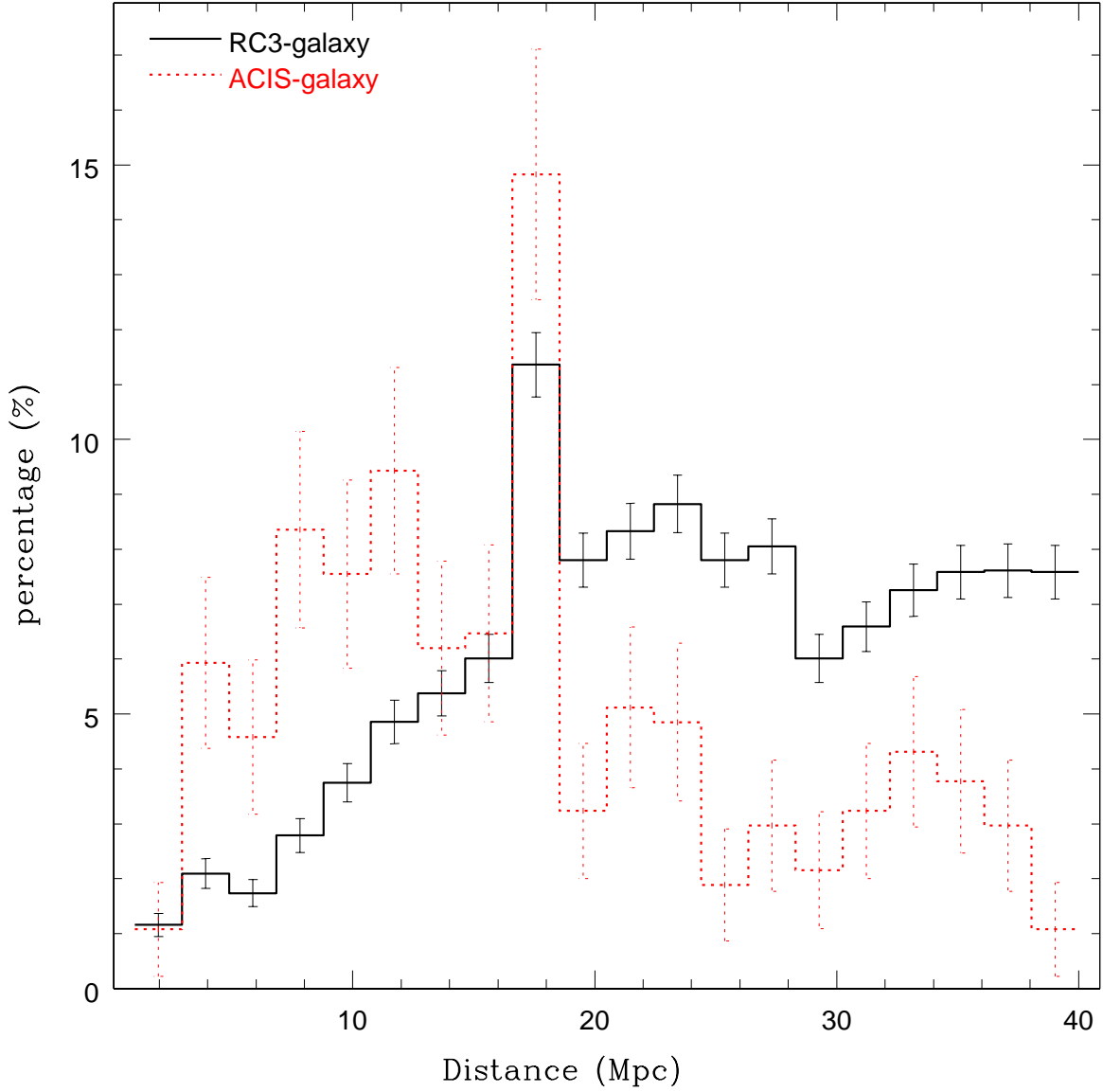}
\caption[Distances for ACIS survey galaxies]{Galaxy distance distribution of
the Chandra ACIS survey galaxy sample in comparison with the RC3 galaxy sample.
More nearby galaxies ($<15$Mpc) were surveyed than those distant galaxies
($>15$ Mpc). }

\end{figure}


\begin{figure}
\plotone{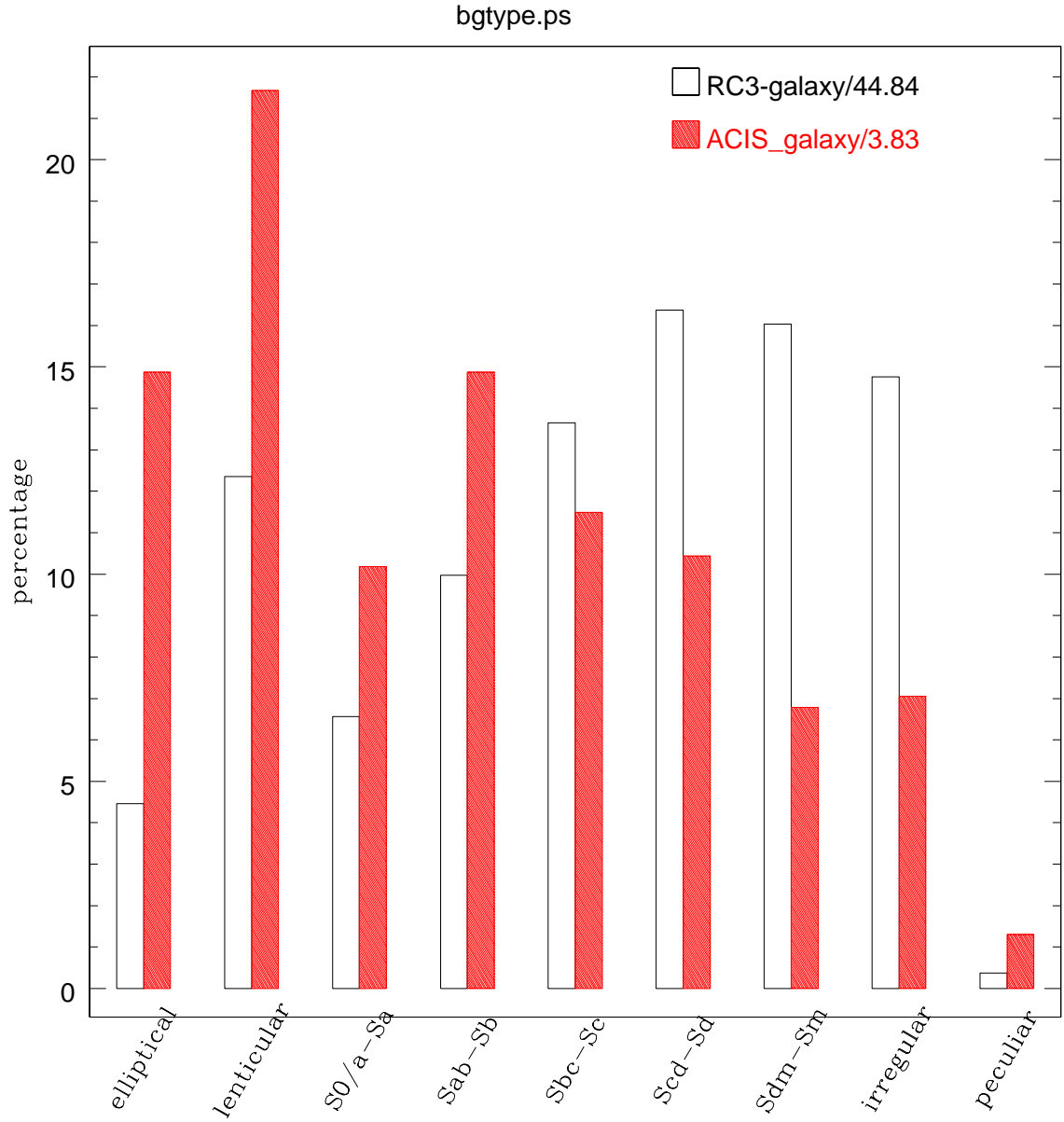}
\caption[Morphological types for ACIS survey galaxies]{The distribution of
morphological types of the Chandra ACIS survey galaxy sample in comparison with
the RC3 galaxy sample.  Relatively more early-types galaxies and Sa--Sb spiral
galaxies were surveyed than other types, since they were the ones observed
most. }

\end{figure}

\begin{figure}
\plotone{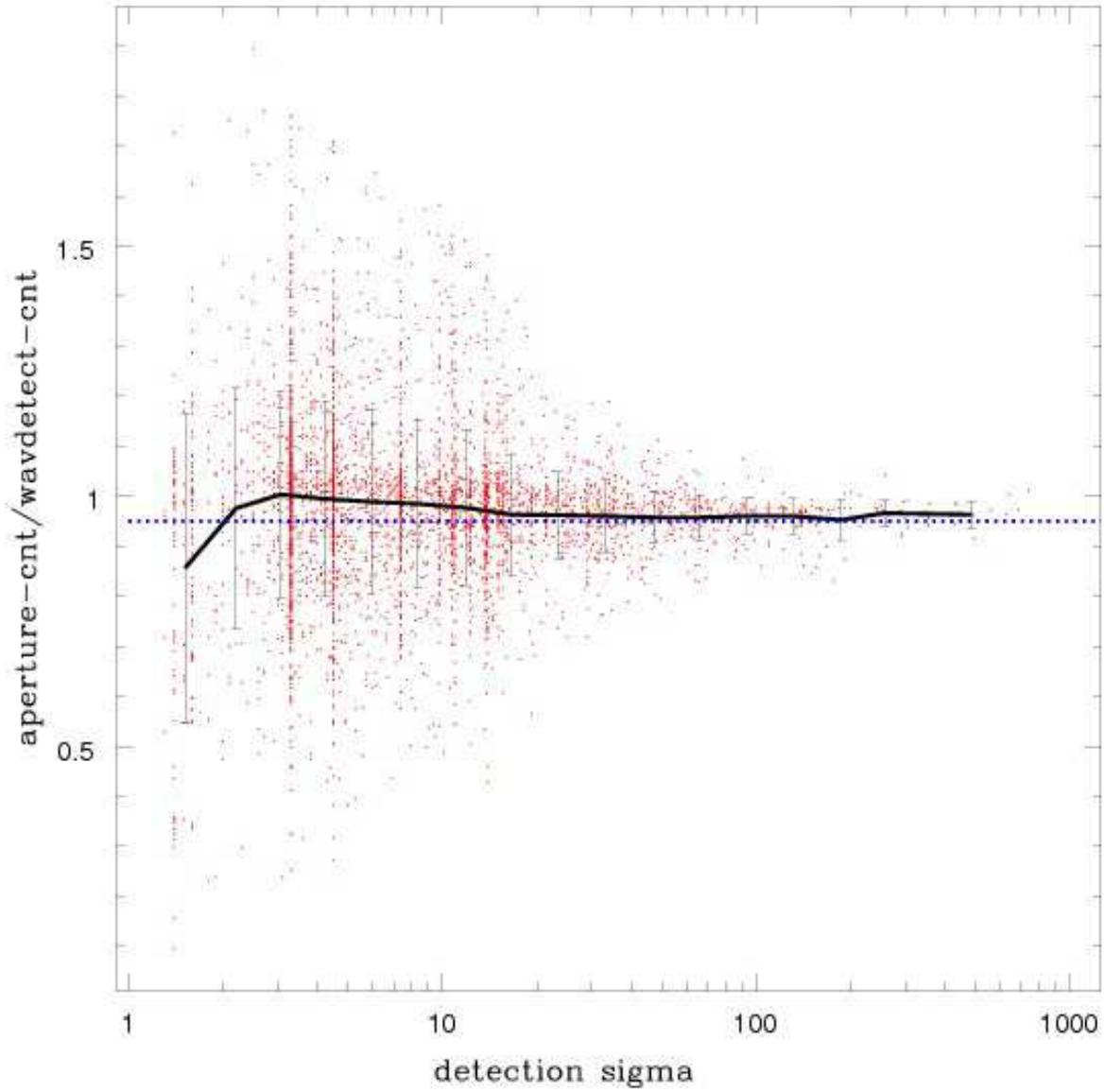}
\caption{Comparison of the photon counts from {\tt wavdetect} and from aperture
photometry that uses a source ellipse enclosing 95\% of the total counts. The
ratio is calculated for 31,400 sources detected in the ACIS survey. The blue
dotted line indicates a ratio of 95\%. The black solid line and the error bars
indicate the average ratio and dispersion for detection sigma intervals. }

\end{figure}

\begin{figure}
\plotone{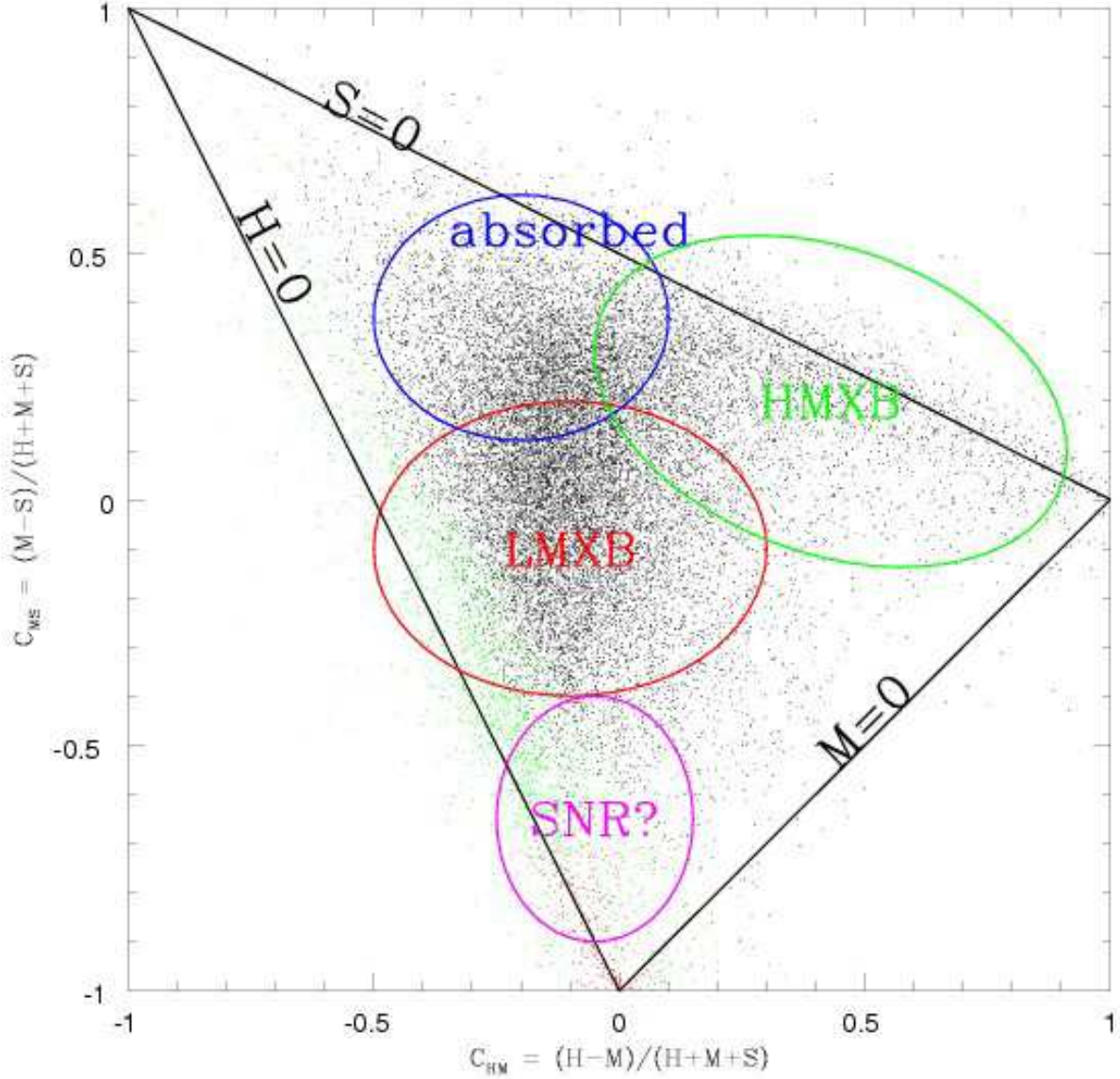}
\caption{The X-ray color-color diagram for sources above 10 counts. Most of the
sources fall within the triangle whose sides stand for zero counts in the soft
(S:0.3-1 keV), medium (M:1-2 keV), and hard (H:2-8 keV) bands. The four regions
located by SNR(?), LMXB, HMXB and absorbed sources are taken from Prestwich et
al.  (2003). We put a question mark after SNR because some sources are
apparently not SNRs as suggested by their variabilities.  The red dots clustered
around $C_{MS} = -1, C_{HM} = 0$ are the supersoft sources, and the blue dots
along H=0 are the quasi-soft sources. }

\end{figure}

\begin{figure}
\plotone{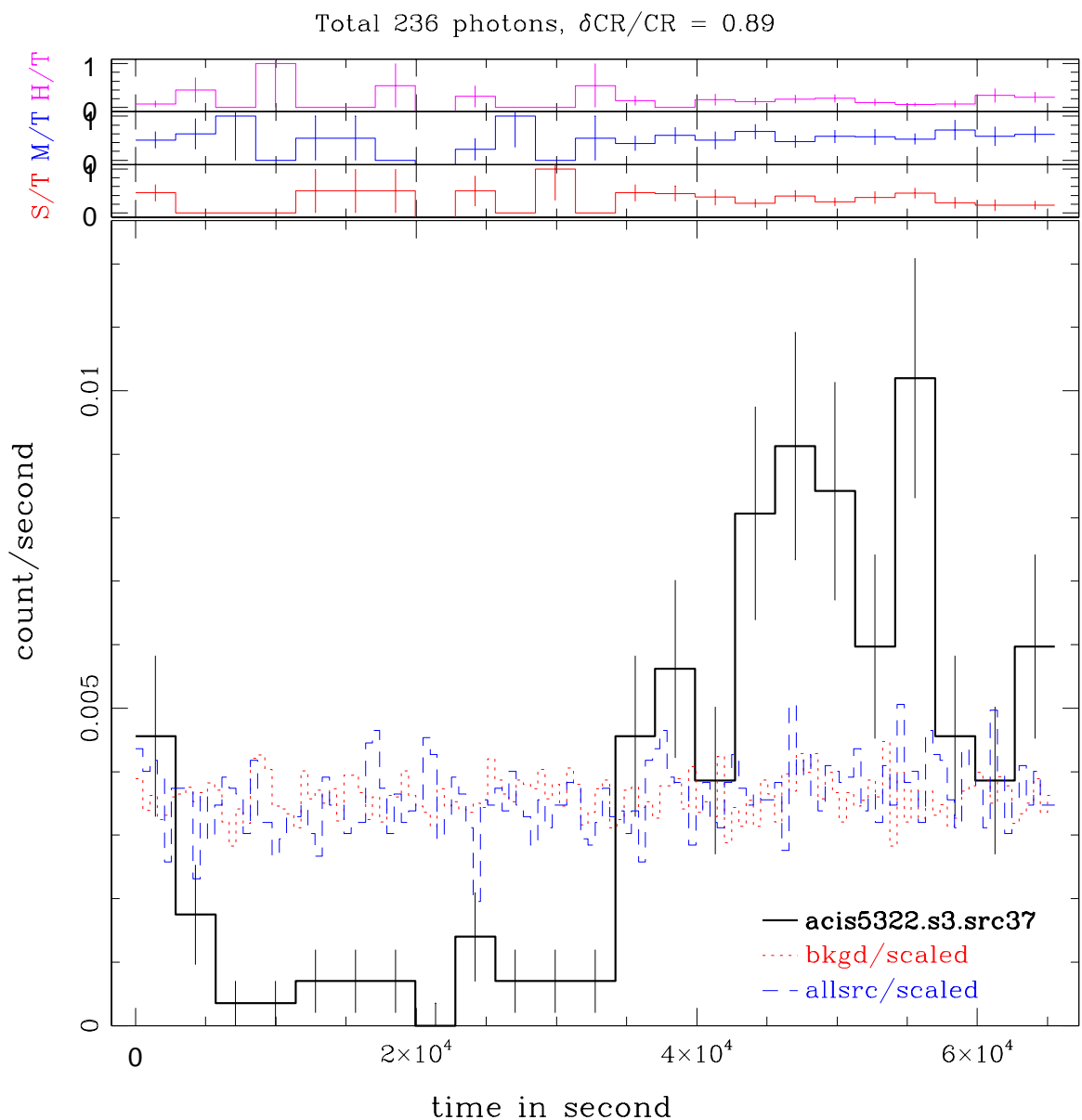}
\caption{The binned light curve for an X-ray source (CXOJ140336.007+541924.74)
in M101 as observed in ObsID 5322. This source (acis5322.s3.src37) exhibited an
X-ray eclipse during the 65 ksec observation. The binned light curves for all
detected sources and for the total background of S3 chip are overplotted for
comparison, which clearly shows that the eclipse is not caused by background or
flares but a behavior unique to this source itself. The three upper panels show
the fractions of photon counts in soft (0.3-1 keV), medium (1-2 keV) and hard
(2-8 keV) bands.  }

\end{figure}

\begin{figure}
\plotone{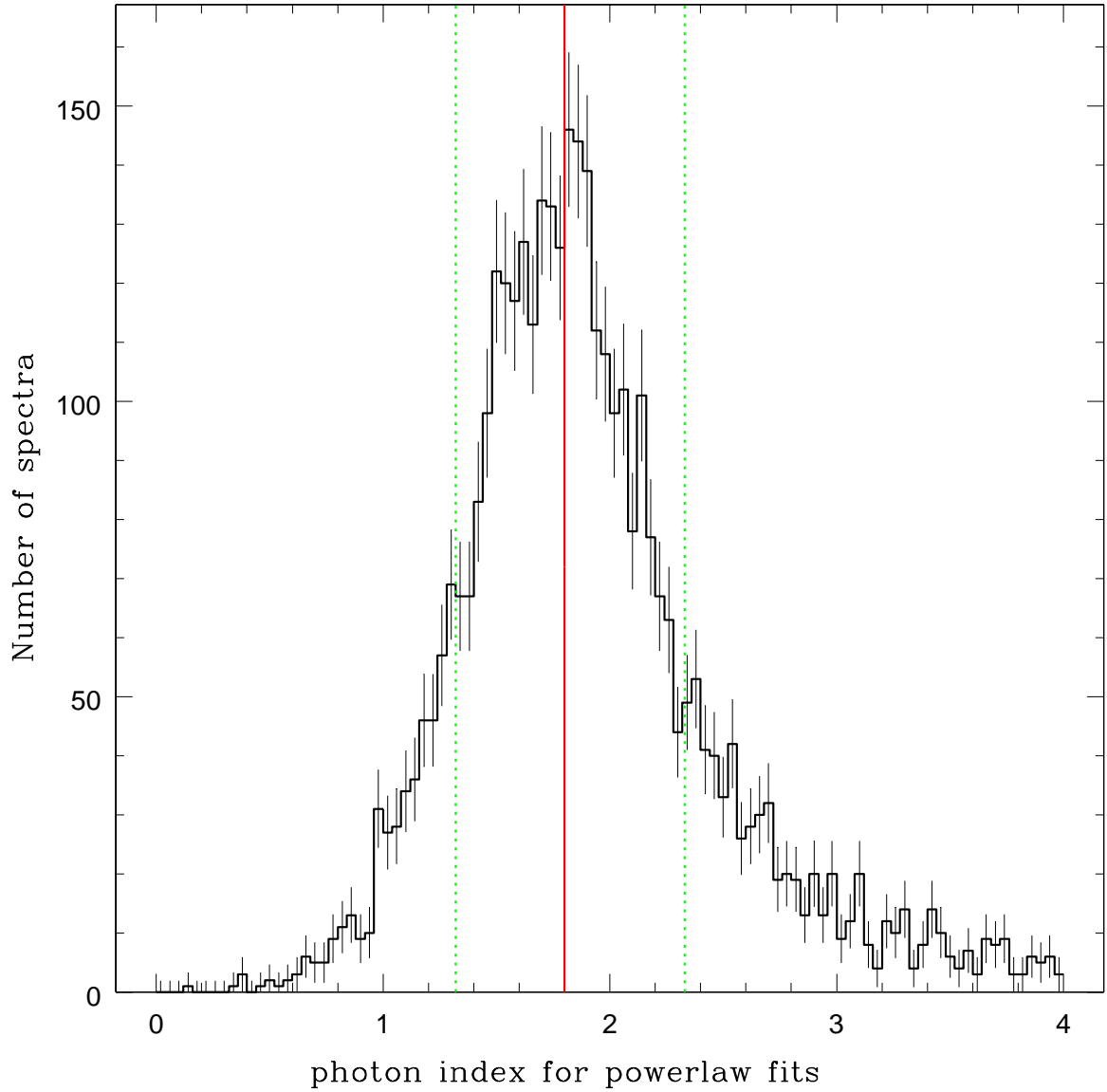}
\caption{Histogram of the photon indices for acceptible power-law fits
($\chi^2_\nu < 1.5$) to $\sim$3900 ACIS spectra from this survey. The
distribution peaks at $\sim1.8$ as indicated by the vertical solid line, with
68.3\% of all photon indices enclosed between 1.3 and 2.3 as indicated by the
vertical dashed lines.}

\end{figure}

\begin{figure}
\plotone{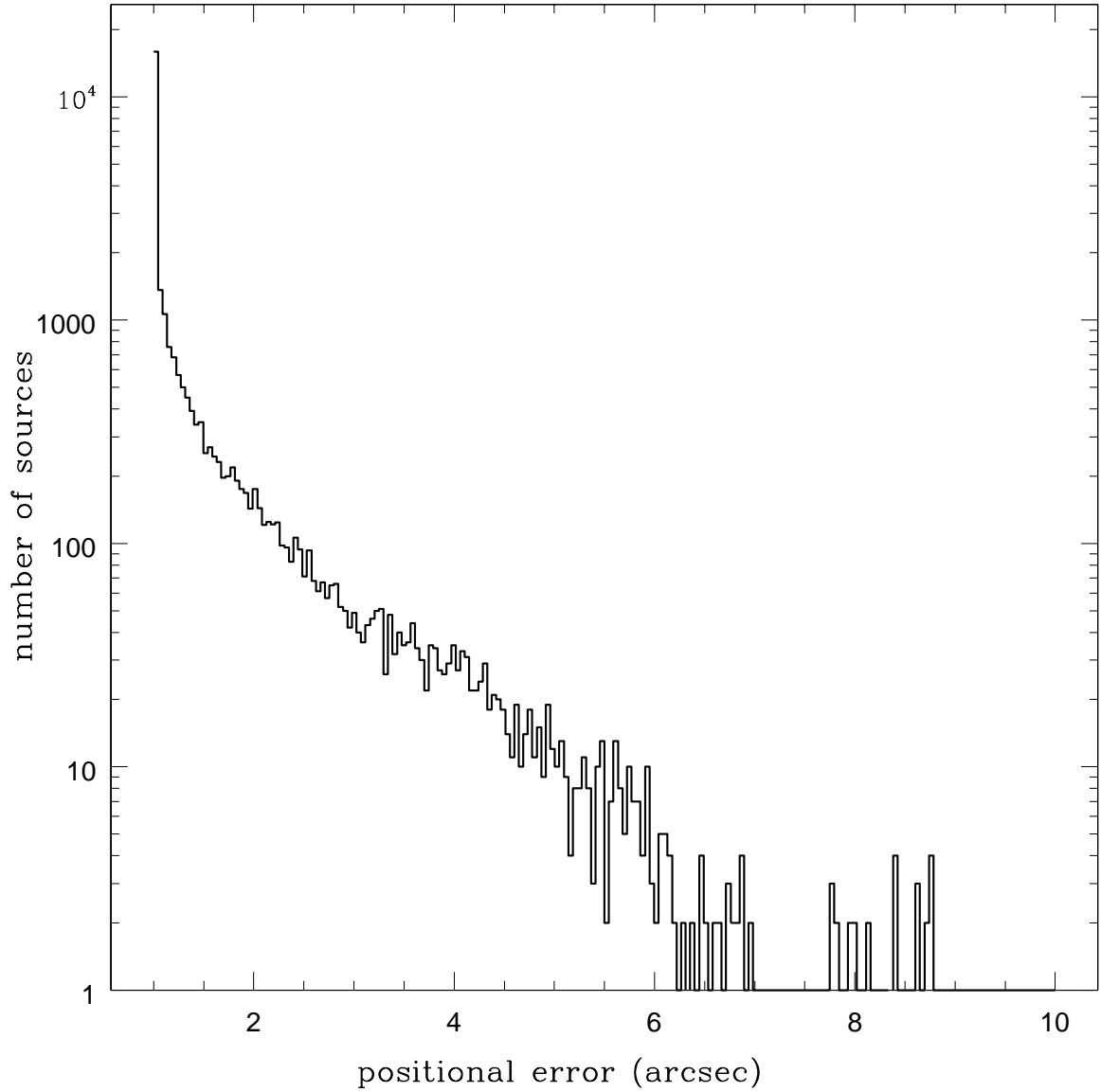}
\caption{Histogram of the positional errors for the 28099 X-ray point sources
detected in 626 ACIS observations analyzed in this survey. The positional
errors are computed following the Kim et al. (2004a) scheme with a conservative
minimum of $1^{\prime\prime}$.  The positional errors are less than
$2^{\prime\prime}$/$3^{\prime\prime}$/$4^{\prime\prime}$/$5^{\prime\prime}$ for
87\%/95\%/98\%/99\% of these sources. }

\end{figure}

\begin{figure}
\plotone{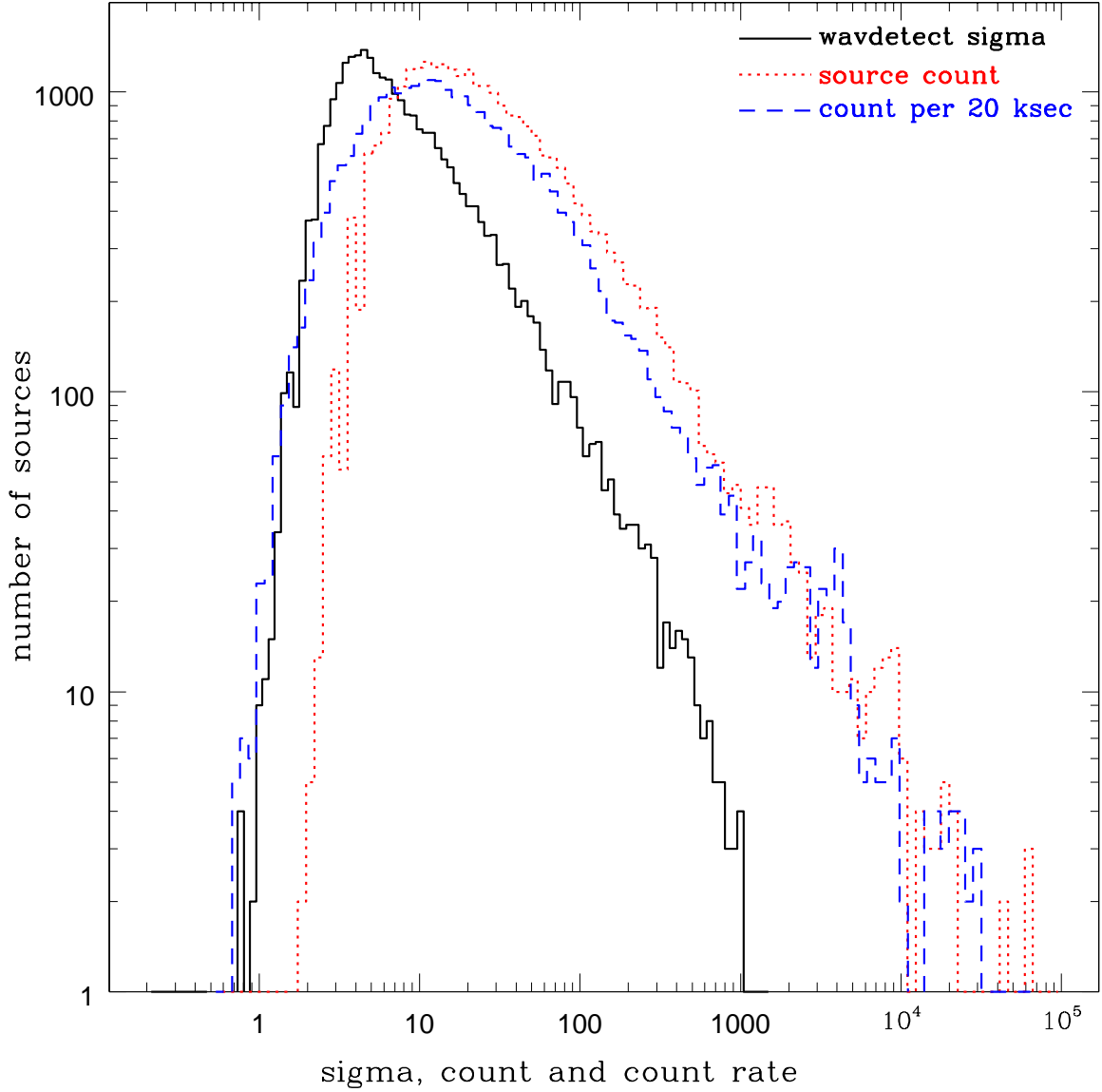}
\caption{Histograms of the wavdetect sigma, source count and count rate for
sources detected in this survey. Of these 28099 sources, about
2.6\%/13.0\%/47.6\%/66.6\% are detected with sigma below 2/3/6/10.  This
composite survey covers a dynamical range of 5 orders of magnitudes in the
source count (from 2 to $10^5$ after background subtraction) and in the count
rate (from $3\times10^{-5}$ to 3 count/sec after vignetting correction). The
source distributions peak at $\sigma\approx4.5$, 14 counts and $5\times10^{-4}$
count/sec. }

\end{figure}

\begin{figure}
\plotone{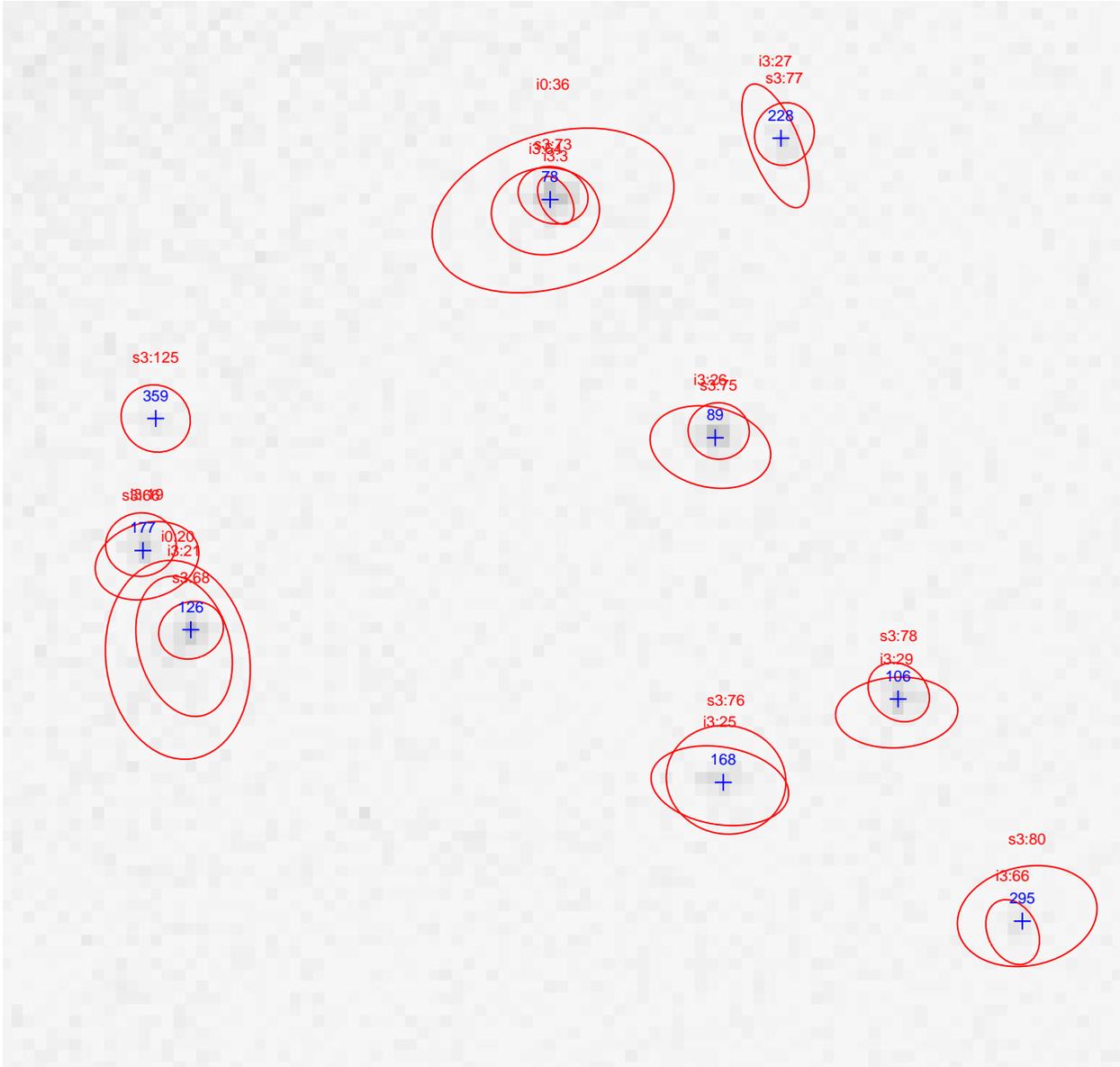}
\caption{Cross identification of sources detected in multiple observations.
The source ellipses as reported by {\tt wavdetect} from individual observations
are overlayed on the merged X-ray image with their labels. Overlapping ellipses
are taken as single sources (crosses). The two close sources to the left of the
image represent a confusing situation where source ellipses from two sources
overlap.  Such confusions can be removed by shrinking the source ellipses, and
single sources are usually identified correctly as in this case. }

\end{figure}

\begin{figure}
\plotone{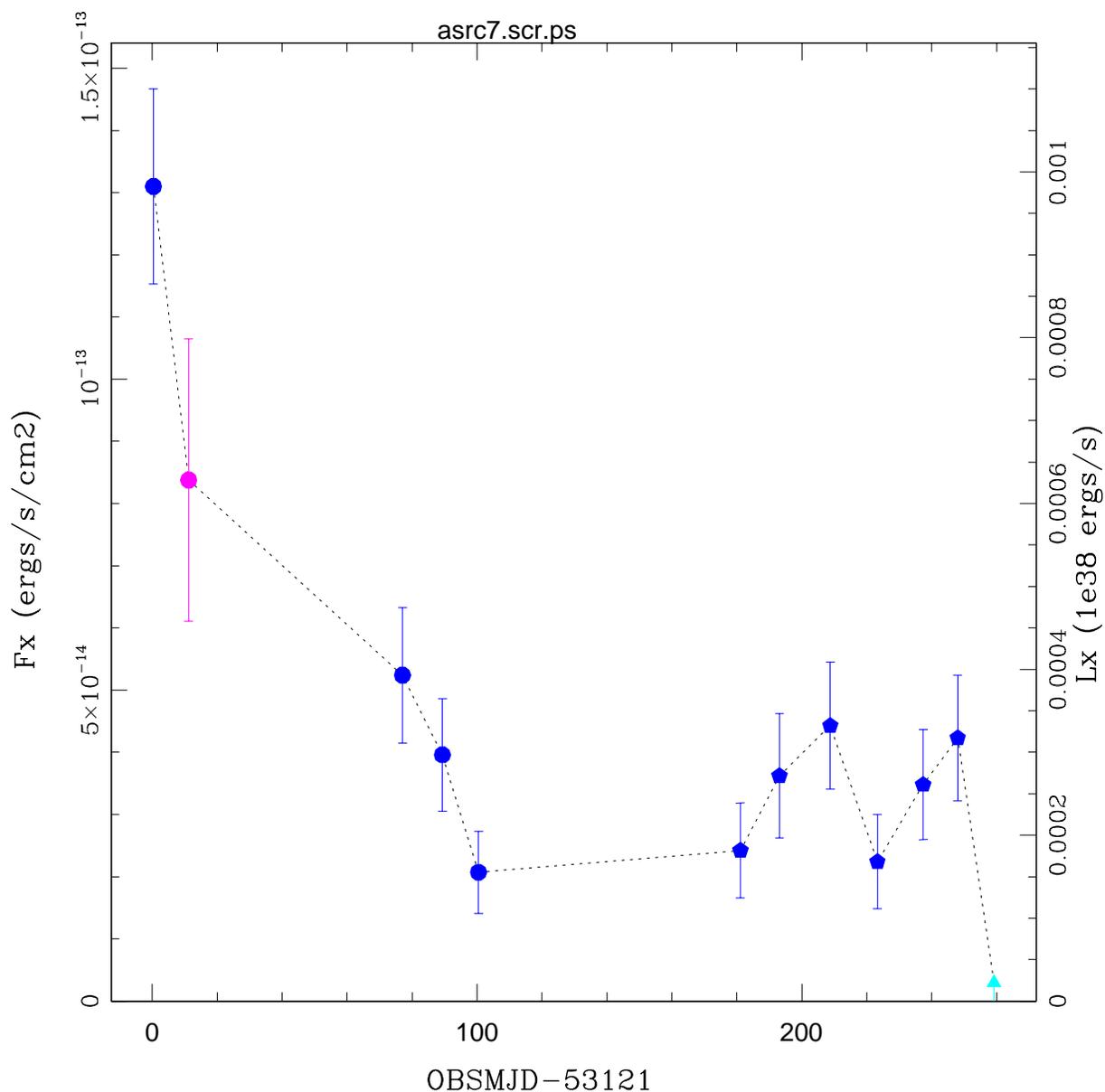}
\caption{The light curve for an X-ray source (CXOJ010029.167-334735.99) in
PGC3589.  The fluxes and luminosities are derived from the count rates for a
$\Gamma=1.7$ powerlaw model in 0.3-8 keV band. The source was observed 12 times
and detected 11 times, with one upper limit (arrow). The point shape indicates
which chip the source was on: circle for S3, pentagon for S2, and square for I
chips.  The point color indicates the QSS/SSS classifications: red for SSS,
purple for QSS, blue for hard, and black for dim sources. }

\end{figure}

\begin{figure}
\plotone{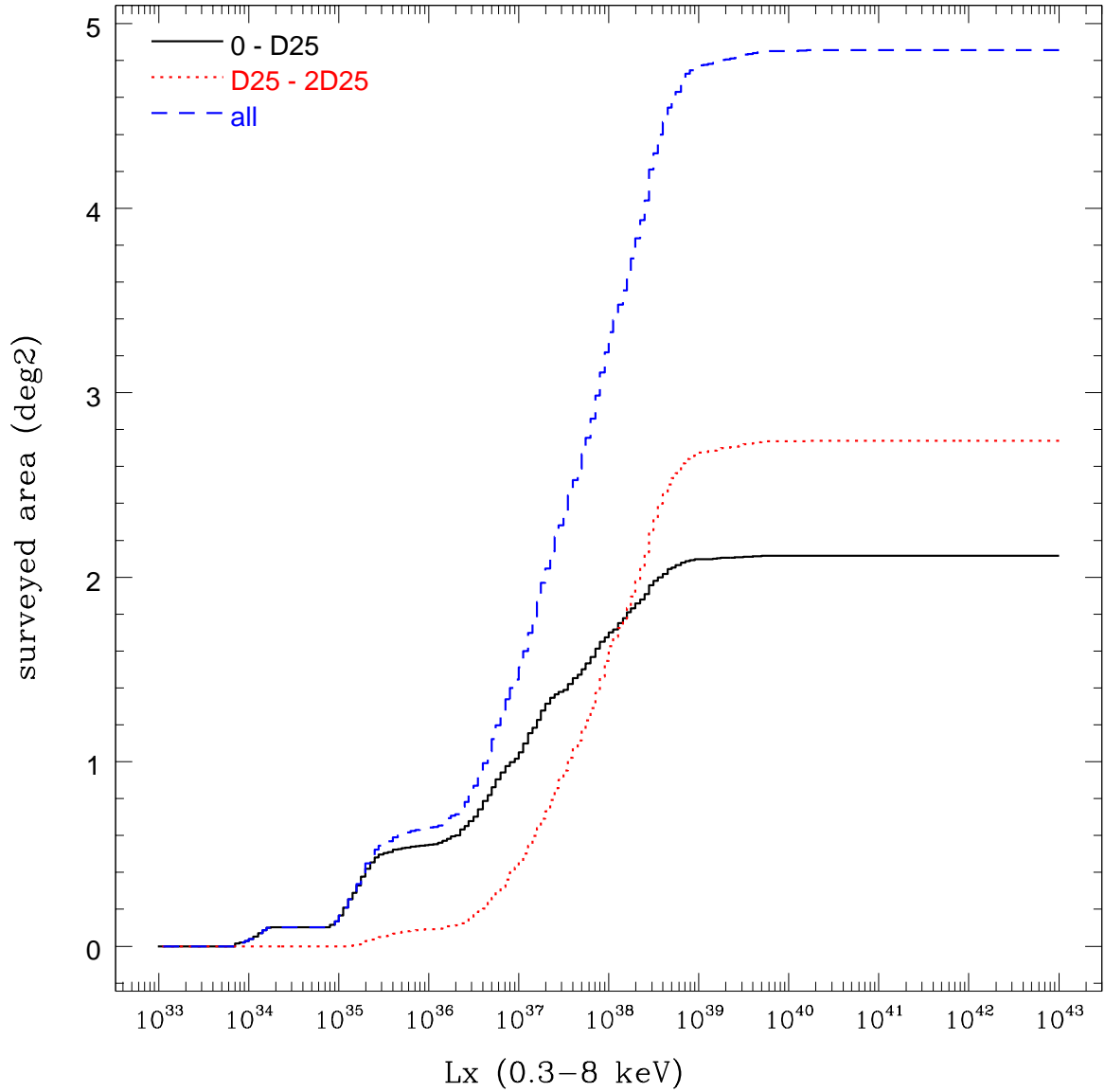}
\caption{Surveyed area curves for the D25 isophotes (solid line), regions
outside D25 but inside 2$\times$D25 isophotes (dotted line), and 2$\times$D25
isophotes of 439 galaxies observed in this ACIS survey.}

\end{figure}

\begin{figure}

\plottwo{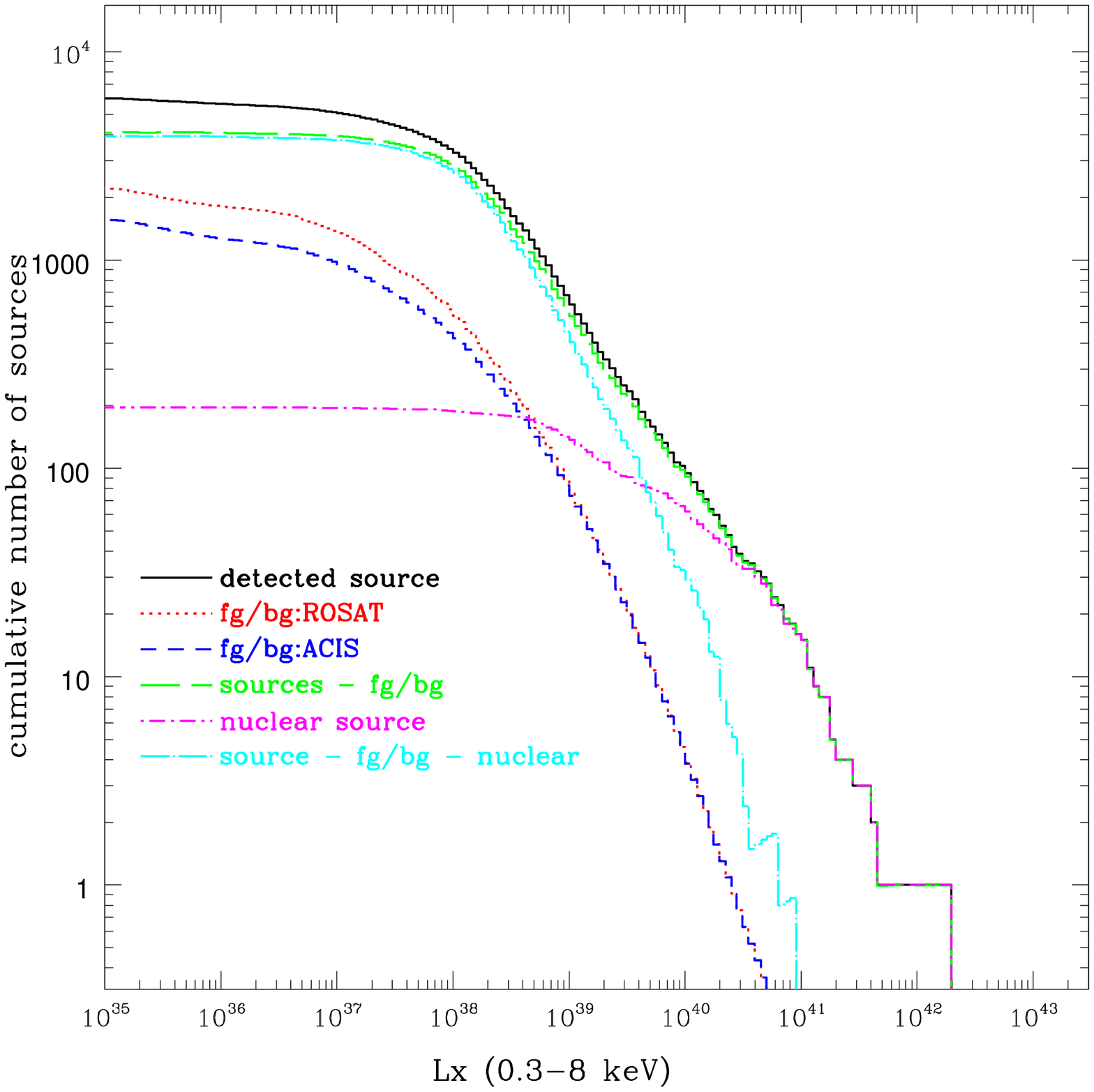}{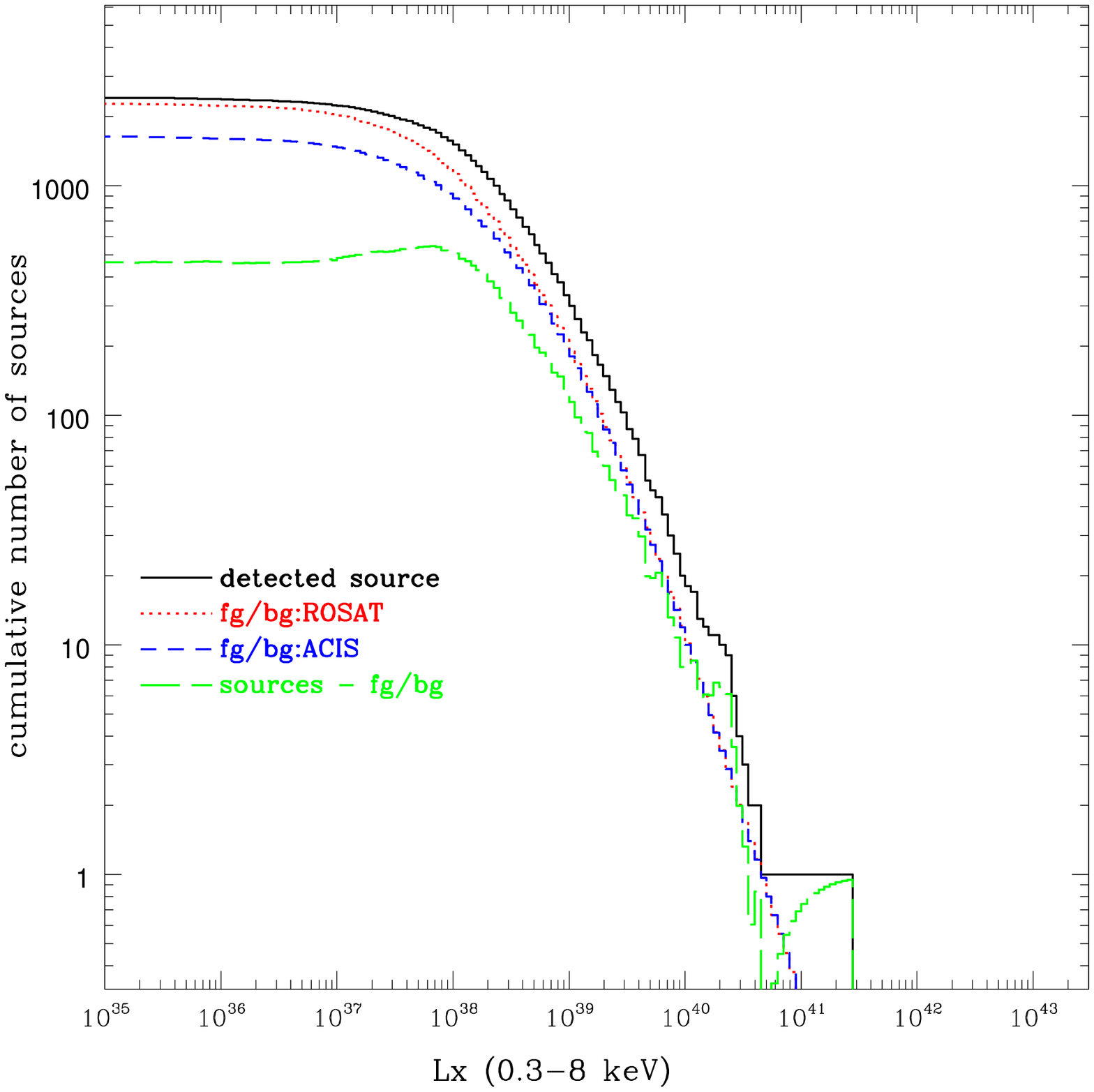} 

\caption{(a) The cumulative numbers of sources within the D25 isophotes of 439
surveyed galaxies.  The various lines represent 
the detected sources (solid line), 
the foreground/background sources predicted with the ROSAT logN-logS relation
(dotted line) 
and the ACIS logN-logS relation (dashed line), 
the ``net'' sources (long dashed line) as subtracting the (average of the two
predictions of) foreground/background sources from the detected sources,
the nuclear sources (dash-dotted line) 
and the ``net'' sources with nuclear sources subtracted (long dash-dotted
line). 
(b) the cumulative numbers of sources outside D25 but inside 2$\times$D25
isophotes. }

\end{figure}



\begin{deluxetable}{lrrlllrlcrrcl}
\rotate
\tablewidth{8in}
\tabletypesize{\scriptsize}
\tablecaption{Properties of surveyed galaxies}
\tablehead{
   \colhead{galaxy} & \colhead{GLON\tablenotemark{a}} & \colhead{GLAT\tablenotemark{b}} & \colhead{R25\tablenotemark{c}} & \colhead{r25\tablenotemark{d}} & \colhead{nH\tablenotemark{e}} & \colhead{Distance\tablenotemark{f}} & \colhead{Morphol.\tablenotemark{g}} & \colhead{Hubble\tablenotemark{h}} & \colhead{$L_B$\tablenotemark{i}} & \colhead{$f_\nu(60\mu m)$\tablenotemark{j}} & \colhead{Group} & \colhead{Notes\tablenotemark{k}} \\
   \colhead{      } & \colhead{(deg)} & \colhead{(deg)} & \colhead{($\prime$)} & \colhead{($\prime$)} & \colhead{(cm$^{-2}$)} & \colhead{(Mpc)} & \colhead{Type} & \colhead{T} & \colhead{($L_\odot$)} & \colhead{(Jy)} & \colhead{No.} & \colhead{} 
}

\startdata

NGC14 & 108.13 & -45.83 & 1.41 & 1.05 & 4.05E+20 & 12.8[T88] & IB(s)m & 10.0 & 2.4e+09 & 0.053 & 256 & \nodata \\
NGC45 &  55.89 & -80.67 & 4.26 & 2.95 & 2.24E+20 & 8.1[T92] & SA(s)dm &  8.0 & 3.8e+09 & 0.012 & 38 & \nodata \\
NGC55 & 332.67 & -75.74 & 16.18 & 2.81 & 1.72E+20 & 1.48[MM] & SB(s)m &  9.0 & 3.0e+09 & 0.076 & 56 & \nodata \\
IC10 & 118.97 &  -3.34 & 3.15 & 2.56 & 5.28E+21 & 0.825[MM] & IBm & 10.0 & 5.4e+08 & $<$0.0002 & 34 & \nodata \\
NGC253 &  97.36 & -87.96 & 13.77 & 3.38 & 1.42E+20 & 3.0[T92] & SAB(s)c &  5.0 & 2.0e+10 & 3.8 & 26 &  HII Sbrst \\
NGC278 & 123.04 & -15.32 & 1.04 & 0.99 & 1.33E+21 & 11.8[T88] & SAB(rs)b &  3.0 & 1.2e+10 & 1.5 & 54 & \nodata \\
PGC3589 & 287.53 & -83.16 & 19.91 & 15.46 & 1.98E+20 & 0.079[MM] & E & \nodata & 1.3e+06 & 7.6e-07 & 3 & \nodata \\
IC1613 & 129.79 & -60.56 & 8.11 & 7.23 & 2.97E+20 & 0.7047[KP] & IB(s)m & 10.0 & 9.0e+07 & 0.00022 & 114 & \nodata \\
\enddata
\end{deluxetable}

\begin{deluxetable}{clrrrcrrrr}
\tablewidth{7in}
\tabletypesize{\scriptsize}
\tablecaption{Groups of surveyed galaxies and ACIS observations}
\tablehead{
   \colhead{Group} & \colhead{galaxies} & \colhead{Number} & \colhead{Exposure} & \colhead{ObsID} & \colhead{Observation} & \colhead{OBSMJD} & \colhead{Exposure} & \colhead{Chips} & \colhead{Number}  \\
   \colhead{No.} & \colhead{      } & \colhead{of Obs} & \colhead{Total (sec)} & \colhead{} & \colhead{Date} & \colhead{} & \colhead{(sec)} & \colhead{used} & \colhead{of sources} 
}

\startdata

  1 & NGC598 & 28 & 1572990 & 787  & 2000-01-11 &  51554.024  & 9657 & 6 7 & 5 \\
    &         &    &         & 1730  & 2000-07-12 &  51737.354  & 50063 & 0 1 2 3 & 73 \\
    &         &    &         & 786  & 2000-08-30 &  51786.010  & 46854 & 6 7 & 46 \\
    &         &    &         & 2023  & 2001-07-06 &  52096.963  & 89942 & 0 1 2 3 & 116 \\
    &         &    &         & 6378  & 2005-09-21 &  53634.011  & 111647 & 0 1 2 3 & 124 \\
    &         &    &         & 6380  & 2005-09-23 &  53636.839  & 90515 & 0 1 2 3 & 122 \\
    &         &    &         & 7170  & 2005-09-26 &  53639.275  & 41487 & 0 1 2 3 & 87 \\
    &         &    &         & 7171  & 2005-09-29 &  53642.191  & 38175 & 0 1 2 3 & 76 \\
\enddata

\end{deluxetable}

\begin{deluxetable}{ccclccrrrrcrrlll}
\rotate
\tablewidth{9in}
\tabletypesize{\tiny}
\tablecaption{X-ray Point sources in nearby galaxies}
\tablehead{
   \colhead{Group} & \colhead{CXOJ\tablenotemark{a}} & \colhead{Pos.\tablenotemark{b}} & \colhead{source\tablenotemark{c}} & \colhead{$\alpha$\tablenotemark{d}} & \colhead{$r_{25}$\tablenotemark{e}} & \colhead{Dis.\tablenotemark{f}} & \colhead{V/D\tablenotemark{g}} & \colhead{$L_{Xmax}$\tablenotemark{h}} & \colhead{$F_{Xavg}$\tablenotemark{i}} & \colhead{$F/F$\tablenotemark{j}} & \colhead{$\sigma_{max}$\tablenotemark{k}} & \colhead{$C_{max}$\tablenotemark{l}} & \colhead{SQH\tablenotemark{m}} & \colhead{Var/KS\tablenotemark{n}} & \colhead{Notes\tablenotemark{o}} \\
   \colhead{No.} & \colhead{} & \colhead{Err.} & \colhead{Name} & \colhead{($\prime$)} & \colhead{} & \colhead{(Mpc)} & \colhead{} & \colhead{(erg/s)} & \colhead{(cgs)} & \colhead{} & \colhead{} & \colhead{} & \colhead{} & \colhead{} & \colhead{} 
}

\startdata

5 & CXOJ132927.545+582534.18 & 1.28 & NGC5204-X2 & 1.264 & 0.81 & 4.3 & 4/4 & 3.2e+38 & 8.95e-14 & 6.55 & 29.777 & 79.048 & h4  &  &   \\
5 & CXOJ132935.609+582408.68 & 1 & NGC5204-X9 & 0.928 & 0.37 & 4.3 & 1/1 & 1.11e+37 & 5.01e-15 & 1 & 3.155 & 6.575 & d1  &  &   \\
5 & CXOJ132935.863+582626.48 & 1.58 & NGC5204-X10 & 1.376 & 0.56 & 4.3 & 7/1 & 9.76e+36 & 4.40e-15 & 2.54 & 5.668 & 23.666 & h1  &  &   \\
5 & CXOJ132936.449+582313.74 & 1 & NGC5204-X11 & 1.838 & 0.74 & 4.3 & 2/1 & 7.52e+36 & 3.39e-15 & 1.58 & 6.803 & 15.219 & h1  &  &   \\
5 & CXOJ132938.405+582342.36 & 1 & NGC5204-X12 & 1.386 & 0.59 & 4.3 & 4/1 & 6.88e+36 & 3.10e-15 & 4.06 & 6.112 & 17.362 & h1  &  &   \\
5 & CXOJ132938.618+582505.59 & 1 & NGC5204-X1 & 0.291 & 0.19 & 4.3 & 13/13 & 9.81e+39 & 2.51e-12 & 4.54 & 640.539 & 9903.551 & q2 h11  & v/1 &   1ULX \\
5 & CXOJ132939.252+582531.08 & 1 & NGC5204-X4 & 0.585 & 0.29 & 4.3 & 13/4 & 4.26e+37 & 1.22e-14 & 14.12 & 19.976 & 100.501 & d3 h1  & v/1 &   \\
5 & CXOJ132940.219+582453.96 & 1 & NGC5204-X3 & 0.527 & 0.34 & 4.3 & 13/6 & 6.39e+37 & 1.71e-14 & 14.92 & 14.946 & 34.106 & q1 d3 h2  &  & *B0Ibp \\
\enddata

\tablenotetext{a}{source names composed from their positions following CXO naming conventions.}
\tablenotetext{b}{source positional uncertainties in arcsec computed from the scheme of Kim et al. (2004). }
\tablenotetext{c}{galactic source names numbered sequentially by its maximum detection significance.}
\tablenotetext{d}{galactic source nuclear separation in arcminutes.}
\tablenotetext{e}{galactic source nuclear separation in unit of the elliptical radius.}
\tablenotetext{f}{distance in Mpc for the source.}
\tablenotetext{g}{numbers the source was observed/detected.}
\tablenotetext{h}{maximum luminosity in erg s$^{-1}$ computed from detections.}
\tablenotetext{i}{average flux in erg s$^{-1}$ cm$^{-2}$ computed from detections.}
\tablenotetext{j}{$F_{max}/F_{min}$ ratio as an extreme variability indicator. }
\tablenotetext{k}{the maximum detection significance.}
\tablenotetext{l}{the maximum photon counts from detections.}
\tablenotetext{m}{statistics for supersoft/quasisoft/hard/dim phases. For
example, s2 means supersoft in two observations, q3 means quasisoft in three
observations, h4 means hard in four observations, and d1 means too dim (below
10 counts) in one observations.}
\tablenotetext{n}{statistics for variability within single observations by the
K-S criterion. For example, v/3 means the source was variable by the K-S
criterion (i.e., $P_{KS}<0.01$) in three observations. }
\tablenotetext{o}{source identification. ULX: ultraluminous X-ray sources, ULS: ultraluminous supersoft sources,
(NGC3031): the galactic nucleus for NGC1313. }

\end{deluxetable}

\begin{deluxetable}{cclrrrrccrcccccc}
\rotate
\tablewidth{9in}
\tabletypesize{\tiny}
\tablecaption{Individual observations of X-ray sources}
\tablehead{
   \colhead{Group} & \colhead{CXOJ\tablenotemark{a}} & \colhead{individual\tablenotemark{b}} & \colhead{Expos.\tablenotemark{c}} & \colhead{OBSMJD\tablenotemark{d}} & \colhead{OAA\tablenotemark{e}} & \colhead{VigF\tablenotemark{f}} & \colhead{St. PosErr\tablenotemark{g}} & \colhead{PosErr\tablenotemark{h}} & \colhead{$\sigma$\tablenotemark{i}} & \colhead{Counts\tablenotemark{j}} & \colhead{$C_{MS}$\tablenotemark{k}} & \colhead{$C_{HM}$\tablenotemark{l}} & \colhead{$F_X$\tablenotemark{m}} & \colhead{$P_{KS}$\tablenotemark{n}} & \colhead{SQH\tablenotemark{o}} \\
   \colhead{No.} & \colhead{} & \colhead{detection} & \colhead{(sec)} & \colhead{} & \colhead{($^{\prime\prime}$)} & \colhead{} & \colhead{($^{\prime\prime},^{\prime\prime}$)} & \colhead{($^{\prime\prime}$)} & \colhead{} & \colhead{} & \colhead{} & \colhead{} & \colhead{(cgs)} & \colhead{} & \colhead{} 
}

\startdata

 5 & CXOJ132927.545+582534.18 & 3940.s3.4 (1) & 5125 & 52887.675 & 343 & 0.938 & 0.83,0.26 & 1.6 & 5.466 & 12.2(3.6) & 0.11(0.74) & -0.15(0.77) & 2.20e-14 & 0.17 & hard \\
  &   & 3941.s3.m0 (1) & 5163 & 52896.487 & 348 & 0.938 & -,- & 1.3 & 29.777 & 79.0(9.9) & 0.13(0.21) & -0.31(0.23) & 1.44e-13 & 0.43 & hard \\
  &   & 3942.s3.3 (1) & 5512 & 52905.398 & 350 & 0.936 & 0.35,0.14 & 1.5 & 14.584 & 49.9(7.4) & 0.10(0.25) & -0.22(0.27) & 9.29e-14 & 0.0017 & hard \\
  &   & 3943.s3.3 (1) & 5183 & 52915.980 & 351 & 0.936 & 0.35,0.14 & 1.5 & 17.322 & 51.0(7.3) & 0.22(0.23) & -0.29(0.23) & 9.93e-14 & 0.21 & hard \\
 5 & CXOJ132935.609+582408.68 & 2028.s3.3 (1) & 10145 & 51918.956 & 96 & 0.992 & 0.15,0.22 & 1 & 3.155 & 6.6(2.6) & -0.67(0.67) & -0.04(0.98) & 5.01e-15 & 0.084 & dim \\
 5 & CXOJ132935.863+582626.48 & 3933.s3.6 (1) & 48937 & 52857.249 & 343 & 0.939 & 0.45,0.26 & 1.6 & 5.668 & 23.7(5.7) & 0.63(0.47) & -0.15(0.46) & 4.40e-15 & 0.49 & hard \\
  &   & 3934.s3.u & 5015 & 52860.400 & 344 & 0.938 & -,- & - & -1.03 & 4.8(4.7) & - & - & 8.93e-15 & - & - \\
  &   & 3935.s3.u & 4764 & 52862.799 & 343 & 0.938 & -,- & - & -0.98 & 4.4(4.5) & - & - & 8.98e-15 & - & - \\
\enddata
\end{deluxetable}

\begin{deluxetable}{ccclccrrrrcrrlll}
\rotate
\tablewidth{9in}
\tabletypesize{\tiny}
\tablecaption{Ultraluminous X-ray sources in nearby galaxies}

\tablehead{
   \colhead{Group} & \colhead{CXOJ\tablenotemark{a}} & \colhead{Pos.\tablenotemark{b}} & \colhead{source\tablenotemark{c}} & \colhead{$\alpha$\tablenotemark{d}} & \colhead{$r_{25}$\tablenotemark{e}} & \colhead{Dis.\tablenotemark{f}} & \colhead{V/D\tablenotemark{g}} & \colhead{$L_{Xmax}$\tablenotemark{h}} & \colhead{$F_{Xavg}$\tablenotemark{i}} & \colhead{$F/F$\tablenotemark{j}} & \colhead{$\sigma_{max}$\tablenotemark{k}} & \colhead{$C_{max}$\tablenotemark{l}} & \colhead{SQH\tablenotemark{m}} & \colhead{Var/KS\tablenotemark{n}} & \colhead{Notes\tablenotemark{o}} \\
   \colhead{No.} & \colhead{} & \colhead{Err.} & \colhead{Name} & \colhead{($\prime$)} & \colhead{} & \colhead{(Mpc)} & \colhead{} & \colhead{(erg/s)} & \colhead{(cgs)} & \colhead{} & \colhead{} & \colhead{} & \colhead{} & \colhead{} & \colhead{} 
}

\startdata

2 & CXOJ140229.904+542118.79 & 1 & NGC5457-X4 & 6.219 & 0.46 & 6.855 & 6/6 & 3.16e+39 & 2.28e-13 & 5.01 & 271.634 & 2562.617 & q2 h4  & v/4 & PM* +iv 1ULX \\
2 & CXOJ140303.950+542734.63 & 1 & NGC5457-X2 & 6.776 & 0.47 & 6.855 & 6/6 & 4.36e+39 & 5.77e-13 & 1.89 & 268.481 & 3146.814 & h6  & v/1 &   1ULX \\
2 & CXOJ140314.302+541806.23 & 1 & NGC5457-X5 & 2.825 & 0.20 & 6.855 & 25/18 & 3.05e+39 & 1.30e-13 & 600.67 & 297.040 & 3870.985 & d2 h16  & v/6 &   1ULX \\
2 & CXOJ140332.381+542102.92 & 1 & NGC5457-X3 & 2.900 & 0.22 & 6.855 & 25/21 & 3.82e+39 & 1.17e-13 & 778.16 & 639.966 & 9233.438 & q6 s11 d1 h3  & v/6 &   1ULX \\
2 & CXOJ140414.196+542604.52 & 1 & NGC5457-X1 & 10.357 & 0.76 & 6.855 & 7/7 & 4.58e+39 & 6.97e-13 & 2.11 & 201.227 & 3435.026 & h7  & v/3 &   1ULX \\
4 & CXOJ095532.952+690033.59 & 1 & NGC3031-X3 & 3.441 & 0.31 & 3.42 & 17/17 & 3.47e+39 & 1.56e-12 & 1.98 & 694.748 & 9482.748 & h17  &  &   1ULX \\
4 & CXOJ095753.314+690348.13 & 1.04 & NGC3031-X1 & 12.490 & 1.66 & 3.42 & 2/2 & 5.25e+39 & 3.58e-12 & 1.09 & 292.351 & 1879.141 & h2  &  & +v 2ULX \\
5 & CXOJ132938.618+582505.59 & 1 & NGC5204-X1 & 0.291 & 0.19 & 4.3 & 13/13 & 9.81e+39 & 2.51e-12 & 4.54 & 640.539 & 9903.551 & q2 h11  & v/1 &   1ULX \\
\enddata

\tablenotetext{a}{source names composed from their positions following CXO naming conventions.}
\tablenotetext{b}{source positional uncertainties in arcsec computed from the scheme of Kim et al. (2004). }
\tablenotetext{c}{galactic source names numbered sequentially by its maximum luminosity.}
\tablenotetext{d}{galactic source nuclear separation in arcminutes.}
\tablenotetext{e}{galactic source nuclear separation in unit of the elliptical radius.}
\tablenotetext{f}{distance in Mpc for the source.}
\tablenotetext{g}{numbers the source was observed/detected.}
\tablenotetext{h}{maximum luminosity in erg s$^{-1}$ computed from detections.}
\tablenotetext{i}{average flux in erg s$^{-1}$ cm$^{-2}$ computed from detections.}
\tablenotetext{j}{$F_{max}/F_{min}$ ratio as an extreme variability indicator. }
\tablenotetext{k}{the maximum detection significance.}
\tablenotetext{l}{the maximum photon counts from detections.}
\tablenotetext{m}{statistics for supersoft/quasisoft/hard/dim phases. For
example, s2 means supersoft in two observations, q3 means quasisoft in three
observations, h4 means hard in four observations, and d1 means too dim (below
10 counts) in one observations.}
\tablenotetext{n}{statistics for variability within single observations by the
K-S criterion. For example, v/3 means the source was variable by the K-S
criterion (i.e., $P_{KS}<0.01$) in three observations. }
\tablenotetext{o}{source identification. ULX: ultraluminous X-ray sources, ULS: ultraluminous supersoft sources,
(NGC3031): the galactic nucleus for NGC1313. }

\end{deluxetable}

\end{document}